# A First-Principles Metal-Semiconductor Interaction Study: Aluminum Adsorption on Ga-Rich GaAs(100) (2 × 1) and $\beta$(4 × 2) Surfaces

Michael L. Mayo and Asok K. Ray*

*Physics Department, University of Texas at Arlington, Arlington, Texas 76019*

*email:akr@uta.edu

# ABSTRACT

*Ab initio* self-consistent total energy calculations using second order Møller-Plesset perturbation theory and Hay-Wadt effective core potentials with associated basis sets (HWECP's) for gallium and arsenic have been used to investigate the chemisorption properties of atomic aluminum on the Ga-rich GaAs(100)-(2 × 1) and $\beta$(4 × 2) surfaces. Finite sized hydrogen saturated clusters with the experimental zinc-blende lattice constant of 5.654 Å and the energy optimized Ga dimer bond length of 2.758 Å have been used to model the semiconductor surface. To investigate the effects of the core electrons of aluminum in the adsorption process, we have represented the Al adatom with both HWECP's and an all electron 6-311++G** basis set. Detailed energetics of chemisorption on the (100) surface layer including adsorption beneath the surface layer at an interstitial site are investigated. Chemisorption energies, nearest surface neighbor bond lengths, Mulliken population analysis, and highest occupied molecular orbital-lowest unoccupied molecular orbital (HOMO-LUMO) gaps are reported for all considered sites of chemisorption.

## I. Introduction

The technological applications of GaAs due to its high electron mobility and direct band gap make it an important system for fundamental and applied research. The industry standard for growing GaAs by molecular beam epitaxy (MBE) is the (100) surface. This surface has the highest aerial density of dangling surface bonds, greater than the (110) or (111) surface and consequently, surface reconstruction is facilitated by these bonds. In this work, we extend our previous work on atomic hydrogen, oxygen, and cesium adsorptions on the Ga-rich GaAs(100) surface [1-4] to study aluminum adsorption on the GaAs(100) surface. As is known, metal-semiconductor interfaces have been of major interest to both experimentalists and theoreticians for an understanding of Ohmic and Schottky barrier contacts and also for extensive industrial applications of semiconductor devices [5-12]. Al-GaAs, in particular, is used extensively in the electronics industry in integrated circuitry and optoelectronic devices. This Al-GaAs interface has been the subject of ongoing research for several decades and at room temperature over time, it is believed that Al undergoes an exchange reaction with surface Ga atoms on the GaAs surface. Ludeke *et al.*[11] used Auger spectroscopy to study the interface behavior and crystallographic relationships of aluminum on the GaAs(100)-c(2 × 8) and the Ga-rich (4 × 6) surfaces. They observed notable differences in the degrees of interface reactivity and crystallographic relationships to Al overlayers on the reconstructed surfaces. No replacement reaction was detected for the c(2 × 8) surface at room temperature and only a partial exchange reaction was observed for the (4 × 6) surface. Chen *et al.* [12] studied ultra thin films deposited on GaAs(100) using positron annihilation induced auger electron spectroscopy (PAES) and electron induced auger electron spectroscopy. They found direct evidence that Ga substitutes for Al in the top layer after Al Deposition and that Ga diffuses into the Al overlayer faster than As. The sensitivity of PAES allowed them to directly monitor the time evolution of the changes in the top layer of an Al layer deposited on GaAs.

In this work for chemisorption of Al on the GaAs(100) surface, we investigate the possible adsorption sites and the nature of the GaAs surface upon adsorption as an *initial* attempt towards a better *ab initio* understanding of the interaction of Al with the Ga-rich GaAs(100) surface. To the best of our knowledge, there are *no* experimental results on atomic Al interaction an the Ga-rich GaAs(100) surface. As a continuation of our previous studies [1-4], we present here a detailed study of atomic aluminum adsorption on the GaAs surface represented, as before, by a set of clusters. Our study as reported here is, to the best of our knowledge, the *first ab initio study of atomic Al adsorption on the Ga-rich GaAs(100) surface.* Specifically investigated are the adsorption sites, chemisorption energies, possibilities of charge transfers between the adatom and the Ga and the As atoms and also the highest occupied molecular orbital-lowest unoccupied molecular orbital (HOMO-LUMO) gaps. We first comment on the computational methodology followed by results.

## II. Computational methodology and results

Both the unrestricted Hartree–Fock (UHF) theory and the many-body perturbation theory (MBPT) as used in this work are well documented in the literature [13-18]. Here we present only a basic equation to define some terms. In the MBPT, the energy is given by the linked diagram expansion:

$$\Delta E = E - E_0 = E_1 + E_{corr} = \sum_{n=0}^{\infty} < \Phi_0 \,|\, V[(E_0 - H_0)^{-1} V]^n \,|\, \Phi_0 >_L \qquad (1)$$

where $\Phi_0$ is taken to be the unrestricted Hartree-Fock (UHF) wavefunction, $H_0$ is the sum of one-electron Fock operators, $E_0$ is the sum of the UHF orbital energies and $V = H - H_0$ perturbation, where $H$ is the usual electronic Hamiltonian. The subscript $L$ indicates the limitation to the linked diagrams. Though one can include various categories of infinite-order summations from Eq. (1), the method is usually limited by termination at some order of perturbation theory. In this work, because of severe demands on computational resources, we have carried out complete second-order (UMP2) calculations, which consist of all single and double-excitation terms for both the bare clusters and the chemisorbed systems.

One of the primary considerations involved in *ab initio* calculations is the type of basis set to be used [19]. Basis sets used in *ab initio* molecular-orbital computations usually involve some compromise between computational cost and accuracy. Keeping in mind the tremendous cost of *ab initio* calculations, specifically for large systems like gallium, and arsenic, we have elected to represent them by effective core potentials or pseudopotentials. In particular we have used the Hay-Wadt effective core potentials (HWECP) and associated basis sets for aluminum, gallium, and arsenic atoms [20]. These core potentials are known to provide excellent agreement with all electron results. To further improve the accuracy of our calculations, one *d* function was added to the Hay-Wadt basis sets. The exponents of the *d* functions were chosen to provide minimum energy for the $Ga_2$, $As_2$, and $Al_2$ dimers with the bond lengths fixed at experimental values [21]. The values for the exponents for gallium and arsenic were found to be $d_{Ga}$ = 0.170 and $d_{As}$ = 0.280 respectively. This procedure has been previously used in out studies of alkali adsorptions on the GaAs (110) surface [22-26]. The value of the exponent for aluminum is found to be $d_{Al}$ = 0.218 (Figure 1). For hydrogen, a [2*s*, 1*p*] basis set was used. All computations were carried out using the parallel version of GAUSSIAN 98 [27] on Compaq Alpha ES20 and ES40 parallel supercomputers at the University of Texas at Arlington.

In this work we considered clusters representing two different reconstructed surfaces, namely the (2 × 1) and *β* (4 × 2) surfaces [1, 2]. Five different clusters were constructed (Figure 2), the smallest being the $Ga_4As_4H_{12}$ with two Ga atoms in the first layer and the largest being $Ga_{19}As_{15}H_{39}$ with nine Ga atoms in the first layer. Each cluster was constructed with Ga and As atoms located at the bulk lattice sites given by the zinc-blende structure with an experimental lattice constant of 5.654 Å. Ga atoms terminated the first or the top layer and the second layer was composed of As atoms while the third layer was composed of Ga atoms. The cluster sizes increased in transverse dimensions as well as number of layers, with the maximum number of layers being three. Hydrogen atoms were used to saturate the dangling bonds, except above the surface, at an energy optimized bond length of 1.511 Å. This is in agreement with the work of Nonoyama *et al.* [28] who used a similar approach for constructing $Ga_4As_4H_{12}$ for chemisorption of atomic and molecular hydrogen on the GaAs(100) surface. The dangling bonds above the clean reconstructed Ga-rich surface are then potential sites for chemisorption. Therefore, simple electron counting rules cannot be applied to these clusters since their surface bonds are not saturated. Due to severe demands on computational resources, total energy optimization was carried out only for the smallest cluster, $Ga_4As_4H_{12}$, by allowing dimerization of the surface Ga atoms. From this process, the reconstructed surface Ga-Ga dimer bond length was found to be 2.758 Å. This dimer bond length was then applied to the $Ga_5As_6H_{16}$, $Ga_7As_6H_{16}$, $Ga_7As_6H_{19}$, and $Ga_{19}As_{15}H_{39}$ clusters. Specifically, the $Ga_4As_4H_{12}$, $Ga_5As_6H_{16}$, $Ga_7As_6H_{16}$, and the $Ga_7As_6H_{19}$ clusters represent the (2 × 1) surface and the $Ga_{19}As_{15}H_{39}$ cluster represents the $\beta(4 \times 2)$ surface [1-4]. Different sizes of clusters are used to represent the same surface because of non-uniqueness of a specific cluster to represent a surface and also to study dependence and convergence of cluster properties with respect to cluster sizes. As a comparison, Guo-Ping and Ruda [29] used a surface Ga-Ga dimer bond length of 2.80 Å in a similar *ab initio* cluster study of the adsorption of sulfur on the Ga-rich GaAs(100) surface. They used a $Ga_7As_7H_{20}$ cluster to represent the $(4 \times 2)$, $(4 \times 6)$, and $(2 \times 6)$ surfaces and found that S atoms chemisorbed preferentially on bridge sites.

The total energies and binding energies of all the clusters at the UHF and the UMP2 levels are shown in table 1. The binding energy per atom for a $Ga_xAs_yH_z$ cluster was calculated in the separated atom limit using the following formula,

$$E_b = (x\,E(Ga) + y\,E(As) + z\,E(H) - E(Ga_x\,As_y\,H_z)) / (x + y + z). \qquad (2)$$

We note that the binding energies oscillate with the number of atoms in the clusters at both levels of theory and the binding energies at the UMP2 level of theory are consistently higher than the corresponding energies at the UHF level of theory (fig. 3). It is known that correlation effects typically increase the binding or cohesive energy in a cluster.

In this study, we have also studied basis set effects on the chemisorption process. In addition to the pseudopotential basis set for Al, we have also used an all-electron basis set, namely a 6-311++G** basis set [30-32] for Al. We have considered six adsorption sites of high symmetry, five surface sites and one interstitial site. Figure 4 illustrates the five surface sites yielding the highest chemisorption energies followed by figure 5 showing the $Ga_4As_4H_{12}$ interstitial cage site. All sites were chosen because of their associated point symmetries. The top, bridge, and the interstitial sites were chosen for their $\sigma_V$ inversion symmetry through a plane perpendicular to the surface dimer mid-point. The cave, hollow, and trough sites were chosen for their $C_{4V}$ point rotational symmetry about an axis normal to the (100) plane. In the top site, the adatom is allowed to approach a path directly on top of a Ga atom, whereas in the interstitial site, the adatom migrated inside the cluster. To examine the relative stability of chemisorption at the different sites, the chemisorption energies are calculated from,

$$E_C = E\,(O) + E(Ga_x As_y H_z) - E(O + Ga_x As_y H_z) \quad (3)$$

Thus, a positive chemisorption energy indicates possibilities for chemisorption. For all surface sites, the height of the adatom above the top Ga layer was varied to yield the maximum chemisorption energy (*i.e.* a minimum of the $E_c$ versus d curve, with the sign of the $E_c$ changed). Typically, several data points were generated to get accurate values of the O adatom distance and the chemisorption energy. Figures 6-39 show the $E_c$ versus $d$ curves.

The numerical results at the at the unrestricted MP2 level of theory for both the HWECP's and associated basis sets augmented by a *d* function and the 6-311++G** basis set for aluminum are shown in table 2. We note that all sites with the exception of the $Ga_4As_4H_{12}$ + Al and $Ga_7As_6H_{19}$ + Al interstitial cave sites are potential sites for aluminum chemisorption. The chemisorption energies for the potential adsorption sites range from 1.204eV to 4.838eV with the 6-311++G** basis set for Al and 1.175eV to 4.728eV with the HWECP for the Al. Also, comparing the chemisorption energies of the two basis representations on the Al adatom, we find no pattern that would suggest that either basis selection for Al yields consistently higher or lower chemisorption energies. In general, however different basis selections yield the same predictions as to the ability for the Al adatom to chemisorb at a particular site. The highest chemisorption energies are found at the trough sites for the $Ga_5As_6H_{16}$ + Al and $Ga_{19}As_{15}H_{39}$ + Al clusters while negative chemisorption energies at the cage site for the $Ga_4As_4H_{12}$ + Al and $Ga_7As_6H_{19}$ + Al clusters suggest that chemisorption does not occur at this site. Notable exceptions are the lowest energy chemisorbed sites. The 6-311++G** basis on Al predicts that the $Ga_4As_4H_{12}$ + Al top sites have the lowest chemisorption energy while the HWECP representation for Al predicts that the $Ga_7As_6H_{19}$ + Al top sites yield the lower chemisorption energies. As noted previously, some of the research suggests a surface exchange reaction between Al and Ga at room temperature. Here the primary consideration is whether or not atomic aluminum will in fact chemisorb on the Ga-rich GaAs(100) surface and at which sites chemisorption occurs preferentially.

Table 3 lists the nearest surface neighbor adatom bond lengths for all sites considered in this study. Although there is slight variance at some sites, the geometrical predictions are the same for both the 6-311++G** and the HWECP representations of the Al adatom. The Al-Ga bond length of 3.146 Å at the $Ga_{19}As_{15}H_{39}$ + Al cave site is the largest bond length for both representations and the 6-311++G** representation of the $Ga_{19}As_{15}H_{39}$ + Al bridge site has the smallest Al-Ga bond length of 2.268 Å. Tables 4 and 5 list the effects on the HOMO-LUMO gaps of the GaAs(100) surface due to adsorption of aluminum. With the exception of the $Ga_4As_4H_{12}$ + Al sites, all the HOMO-LUMO gaps increase in value from 0.145eV for Al adsorption on the $Ga_{19}As_{15}H_{39}$ cluster at the cave site to 4.003eV for Al adsorption on the $Ga_7As_6H_{19}$ cluster at the bridge site for the 6-311++G** representation of the Al adatom. The same is true for the HWECP representation of the Al adatom, with the lowest value of 0.246eV for for Al adsorption on the $Ga_{19}As_{15}H_{39}$ cluster at the cave site to 4.359eV for Al adsorption on the $Ga_7As_6H_{19}$ cluster at the bridge site. Our results thus suggest, in general, *a possible transition to* *insulating behavior for the GaAs(100) surface due to aluminum adsorption*. There appears to be no correlation between chemisorption energy and HOMO-LUMO gap for Al adsorption on the GaAs(100) surface. It is interesting to note that though both cage sites yield negative chemisorption energies but the $Ga_4As_4H_{12}$ + Al cage site shows a decrease in HOMO-LUMO gap and the $Ga_7As_6H_{19}$ + Al cage site shows an increase in HOMO-LUMO gap for both 6-311++G**basis and pseudopotential representations of the Al adatom.

We have also carried out an analysis of the atomic charge distributions using Mulliken population analysis [33-36]. While the magnitudes of the charge exchange between the two representations of the Al adatom vary, the general trends between the two treatments are the same. It is worth noting that the cage sites, due to their negative chemisorption energies, are assumed not to be bound. The 6-311++G** treatment of the Al atom predicts that the sites with the largest transfer of charge from the Al adatom are the $Ga_7As_6H_{19}$ + Al cage site with 1.158*e* of its charge transferred followed by the charge transfer from the Al adatom at the $Ga_{19}As_{15}H_{39}$ + Al 5a and 5b trough sites with 0.807*e* and 0.815*e* of charge transfer respectively. These sites are then followed by the $Ga_4As_4H_{12}$ + Al cage site with charge transfer of 0.763*e* from the Al adatom. This is in contrast to the HWECP treatment for the Al adatom where the two unbound cage sites yield the greatest transfer of charge from the Al adatom; 1.025*e* for the $Ga_7As_6H_{19}$ + Al cage site and 0.731*e* for the $Ga_4As_4H_{12}$ + Al cave site followed by the $Ga_{19}As_{15}H_{39}$ + Al 5a and 5b trough sites with 0.625*e* and 0.607*e* charge transfer respectively. It is interesting to note that both the least stable systems and the most stable systems appear to have the greatest charge transfer from the Al adatom. In general, for all considered clusters, the surface Ga atoms tend to gain charge near the site of chemisorption and second layer As atoms tend to gain charge as well. The charge transfer to the third layer is negligible with the exception of the unbound cage where the third layer Ga atoms gain a significant amount of charge.

## III. Conclusions

The effects of Al adsorption on the Ga-rich GaAs(100) surface have been investigated using first-principles quantum mechanical perturbative methods. It was observed that for the six sites of high symmetry considered, all sites, except the cave sites, are candidates for chemisorption of aluminum. The results of two different basis set treatments namely, the all-electron 6-311++G** basis set and the pseudopotential basis set for aluminum have been compared. In general, both representations yielded similar results, neither changing the overall conclusions of this study. The effects of aluminum chemisorption on the HOMO-LUMO gap were studied and, with the exception of the smallest $Ga_4As_4H_{12}$ clusters, the gap increased for all sites considered suggesting a possible transition to insulating behavior of the Ga-rich GaAs(100) surface. Mulliken population analysis indicates that in all cases the Al adatom loses charge. This charge loss is generally gained mostly by surface Ga atoms followed by the second layer As atoms.

## REFERENCES

1. R. Schailey: "*An Ab Initio Cluster Study of Chemisorption of Atomic Cesium and Hydrogen on Reconstructed Surfaces of Gallium Rich Gallium Arsenide (100) Surface",* Ph.D. Dissertation, University of Texas at Arlington, 1999.

2. R. Schailey and A. K. Ray: An *ab initio* cluster study of chemisorption of atomic Cs on Ga-rich GaAs (100) (2×1), (2×2), and $\beta$ (4×2) surfaces. *J. Chem. Phys.* **111**, 8628 (1999).

3. R. Schailey and A. K. Ray: A cluster approach to hydrogen chemisorption on the GaAs (100) surface. *Comp. Mat. Sci.* **22**, 169 (2001).

4. M. L. Mayo and A. K. Ray: (*to be published).*

5. E. J. Mele and J. D. Joannopoulos, Surface-barrier formation for A1 chemisorbed on GaAs (110). *Phys. Rev. Lett.* **42**, 1094 (1979).

6. E. J. Mele and J. D. Joannopoulos: Electronic structure of Al chemisorbed on GaAs (110). *J. Vac. Sci. Tech.* **16**, 1154 (1979).

7. S. J. Eglash, M. D. Williams, P. H. Mahowald, N. Newman, I. Lindau, and W. E. Spicer: Aluminum Schottky barrier formation on arsenic capped and heat cleaned MBE GaAs (100). *J. Vac. Sci. Tech. B* **2**, 481 (1984).

8. J. P. A. Charlesworth, R. W. Godby, and R. J. Needs: First-principles calculations of many-body band-gap narrowing at an Al/GaAs(110) interface. *Phys. Rev. Lett.* **70**, 1685 (1993).

9. S. A. Lourenço, I. F. L. Dias, J. L. Duarte, E. Laureto, F. A. Meneses, J. R. Leite, I. Mazzaro: Temperature dependence of optical transitions in AlGaAs. *J. Appl. Phys.* **89**,

6159 (2001).

10. C. Berthod, N. Binggeli, A. Baldereschi: Schottky barrier heights at polar metal/semiconductor interfaces. *Phys. Rev. B* **68**, 085323 (2003).

11. R. Ludeke and G. Landgren: Interface behavior and crystallographic relationships of aluminum on GaAs (100) surfaces. *J. Vac. Sci. Tech.* **19**, 667 (1981).

12. W.-C. Chen, N. G. Fazleev, A. H. Weiss: Study of ultra-thin Al films deposited on GaAs(100) using positron annihilation induced Auger electron spectroscopy and electron induced Auger electron spectroscopy. *Rad. Phys. Chem.* **68**, 619 (2003).

13. J. Goldstone: Derivation of the Brueckner many-body theory. *Proc. R. Soc. London*, Ser. **A 239**, 267 (1956).

14. P. O. Löwdin: Studies in perturbation theory IX. connection between various approaches in the recent development - evaluation of upper bounds to energy eigenvalues in Schrödinger's perturbation theory. *J. Math. Phys.* **6**, 1341 (1965).

15. P.O. Löwdin: Studies in perturbation theory. X. lower bounds to energy egenvalues in perturbation-theory ground state. *Phys. Rev.* **139**, A357 (1965).

16. R. J. Bartlett: Many-body perturbation theory and coupled cluster theory for electron correlation in molecules. *Ann. Rev. Phys. Chem.* **32**, 359 (1981).

17. W. J. Hehre, L. Random, P. v. R. Schleyer, and J. A. Pople: *Ab initio Molecular Orbital Theory* (Wiley, New York, 1982).

18. A. Szabo and N. S. Ostlund: *Modern Quantum Chemistry* (MacMillian, New York, 1982).

19. E. R. Davidson and D. Feller: Basis set selection for molecular calculations. *Chem.*

*Rev.* **86**, 681 (1986).

20. P. J. Hay and W. R. Wadt: *Ab initio* effective core potentials for molecular calculations: potentials for main group elements Na to Bi. *J. Chem. Phys.* **82**, 284 (1985).

21. D. R. Lide ed.: *CRC Handbook of Chemistry and Physics, 83rd. ed.* (CRC Press, Cleveland, 2002).

22. K. M. Song, "*An Ab Initio Study of Alkali Metal Adsorption on Gallium Arsenide (110) Surface"*, (Ph. D. Dissertation, The University of Texas at Arlington, 1994).

23. K. M. Song, D. C. Khan, and A. K. Ray: Correlation study of sodium-atom chemisorption on the GaAs (110) surface. *Phys Rev. B* **49**, 1818 (1994).

24. K. M. Song and A. K. Ray: *Ab initio* study of cesium chemisorption on the GaAs (110) surface. *Phys Rev. B* **50**, 14255 (1994).

25. K. M. Song and A. K. Ray: An *ab initio* study of potassium chemisorption on the GaAs (110) surface *J. Phys: Cond. Matt.* **6**, 9571 (1994).

26. K. M. Song and A. K. Ray: A cluster study of Rb atom chemisorption on a GaAs (110) surface. *J. Phys.: Cond. Matt.* **8**, 6617 (1996).

27. Gaussian 98 (Revision A.9), M. J. Frisch, G. W. Trucks, H. B. Schlegel, G. E. Scuseria, M. A. Robb, J. R. Cheeseman, V. G. Zakrzewski, J. A. Montgomery, Jr., R. E. Stratmann, J. C. Burant, S. Dapprich, J. M. Millam, A. D. Daniels, K. N. Kudin, M. C. Strain, O. Farkas, J. Tomasi, V. Barone, M. Cossi, R. Cammi, B. Mennucci, C. Pomelli, C. Adamo, S. Clifford, J. Ochterski, G. A. Petersson, P. Y. Ayala, Q. Cui, K. Morokuma, D. K. Malick, A. D. Rabuck, K. Raghavachari, J. B. Foresman, J. Cioslowski, J. V. Ortiz, A. G. Baboul, B. B. Stefanov, G. Liu, A. Liashenko, P.

Piskorz, I. Komaromi, R. Gomperts, R. L. Martin, D. J. Fox, T. Keith, M. A. Al-Laham, C. Y. Peng, A. Nanayakkara, C. Gonzalez, M. Challacombe, P. M. W. Gill, B. G. Johnson, W. Chen, M. W. Wong, J. L. Andres, M. Head-Gordon, E. S. Replogle and J. A. Pople, Gaussian, Inc., Pittsburgh PA, 1998.

28. S. Nonoyama, Y. Aoyagi, and S. Namba: *Ab initio* cluster study of hydrogen with the GaAs (100) surface. *Jap. J. Appl. Phys.* **31**, 1298 (1992).

29. J. Guo-Ping and H. E. Ruda: *Ab initio* studies of S chemisorption on GaAs (100). *J. Appl. Phys.* **79**, 3758 (1996).

30. R. C. Binning Jr. and L. A. Curtiss: Compact contracted basis sets for third-row atoms: Ga-Kr. *J. Comp. Chem.* **11**, 1206 (1990).

31. M. P. McGrath and L. Radom: Extension of Gaussian-1 (G1) theory to bromine-containing molecules. *J. Chem. Phys.* **94**, 511 (1991).

32. L. A. Curtiss, M. P. McGrath, J.-P. Blaudeau, N. E. Davis, R. C. Binning Jr., and L. Radom: Extension of Gaussian-2 theory to molecules containing third-row atoms Ga-Kr. *J. Chem. Phys.* **103**, 6104 (1995).

33. R. S. Mulliken: Electronic population analysis on LCAO-MO molecular wave functions I. *J. Chem. Phys.* **23**, 1833 (1955).

34. R. S. Mulliken: Electronic population analysis on LCAO-MO molecular wave functions II. overlap populations, bond Orders, and covalent bond energies. *J. Chem. Phys.* **23**, 1841(1955).

35. R. S. Mulliken: Electronic population analysis on LCAO-MO molecular wave functions III. effects of hybridization on overlap and gross AO populations. *J. Chem. Phys.* **23,** 2338 (1955).

36. R. S. Mulliken: Electronic population analysis on LCAO-MO molecular wave functions IV. bonding and antibonding in LCAO and valence-bond theories. *J. Chem. Phys.* **23,** 2343 (1955).

Table 1. Total energy (in a.u.) and binding energy (in eV) of the (2 × 1) and $\beta$(4 × 2) bare clusters.

| Cluster | $E_{tot}$(UHF) | $E_{tot}$(UMP2) | $E_b$(UHF) | $E_b$(UMP2) |
|---|---|---|---|---|
| $Ga_4As_4H_{12}$ | -38.653 | -39.355 | 1.444 | 1.944 |
| $Ga_5As_6H_{16}$ | -54.724 | -55.749 | 1.290 | 1.847 |
| $Ga_7As_6H_{16}$ | -58.817 | -59.947 | 1.386 | 1.950 |
| $Ga_7As_6H_{19}$ | -60.576 | -61.713 | 1.481 | 1.998 |
| $Ga_{19}As_{15}H_{39}$ | -149.264 | -152.236 | 1.301 | 1.900 |

Table 2. Chemisorption energy (in eV) vs. cluster size and smmetry.

| Site | Symmetry | Cluster | 6-311++G** $E_C$(UMP2) | H-W $E_C$(UMP2) |
|---|---|---|---|---|
| 1 (Top) | | | | |
| 1a | 2 × 1 | Al + $Ga_4As_4H_{12}$ | 1.204 | 1.626 |
| 1b | 2 × 1 | Al + $Ga_4As_4H_{12}$ | 1.204 | 1.626 |
| 1a | 2 × 1 | Al + $Ga_7As_6H_{19}$ | 1.409 | 1.175 |
| 1b | 2 × 1 | Al + $Ga_7As_6H_{19}$ | 1.848 | 1.331 |
| 1b | 4 × 2 | Al + $Ga_{19}As_{15}H_{39}$ | 2.377 | 2.161 |
| 2 (Bridge) | 2 × 1 | Al + $Ga_4As_4H_{12}$ | 2.327 | 2.245 |
| | 2 × 1 | Al + $Ga_7As_6H_{19}$ | 2.605 | 2.795 |
| | 4 × 2 | Al + $Ga_{19}As_{15}H_{39}$ | 2.765 | 2.841 |
| 3 (Hollow) | 2 × 1 | Al + $Ga_7As_6H_{19}$ | 2.618 | 2.578 |
| | 4 × 2 | Al + $Ga_{19}As_{15}H_{39}$ | 3.470 | 3.167 |
| 4 (Cave) | 2 × 1 | Al + $Ga_7As_6H_{16}$ | 2.572 | 2.454 |
| | 4 × 2 | Al + $Ga_{19}As_{15}H_{39}$ | 2.462 | 2.693 |
| 5 (Trough) | 2 × 1 | Al + $Ga_5As_6H_{16}$ | 4.097 | 3.965 |
| 5a | 4 × 2 | Al + $Ga_{19}As_{15}H_{39}$ | 4.838 | 4.728 |
| 5b | 4 × 2 | Al + $Ga_{19}As_{15}H_{39}$ | 4.659 | 4.556 |
| 6 (Cage) | 2 × 1 | Al + $Ga_4As_4H_{12}$ | -3.689 | -3.994 |
| | 2 × 1 | Al + $Ga_7As_6H_{19}$ | -2.657 | -3.194 |

Table 3. Bond Length (in A) of the Al adatom vs. cluster size and symmetry.

| Sites | Symmetry | Cluster | Adatom-nearest surface neighbor bond length | Adatom-nearest surface neighbor bond length |
|---|---|---|---|---|
| | | | 6-311++G** | H-W |
| 1 (Top) | | | | |
| 1a | 2 × 1 | Al + $Ga_4As_4H_{12}$ | 2.500 | 2.700 |
| 1b | 2 × 1 | Al + $Ga_4As_4H_{12}$ | 2.500 | 2.700 |
| 1a | 2 × 1 | Al + $Ga_7As_6H_{19}$ | 2.900 | 2.700 |
| 1b | 2 × 1 | Al + $Ga_7As_6H_{19}$ | 2.500 | 2.700 |
| 1b | 4 × 2 | Al + $Ga_{19}As_{15}H_{39}$ | 2.300 | 2.500 |
| 2 (Bridge) | 2 × 1 | Al + $Ga_4As_4H_{12}$ | 2.682 | 2.682 |
| | 2 × 1 | Al + $Ga_7As_6H_{19}$ | 2.865 | 2.513 |
| | 4 × 2 | Al + $Ga_{19}As_{15}H_{39}$ | 2.268 | 2.430 |
| 3(Hollow) | 2 × 1 | Al + $Ga_7As_6H_{19}$ | 2.803 | 2.803 |
| | 4 × 2 | Al + $Ga_{19}As_{15}H_{39}$ | 3.018 | 3.122 |
| 4 (Cave) | 2 × 1 | Al + $Ga_7As_6H_{16}$ | 2.908 | 2.908 |
| | 4 × 2 | Al + $Ga_{19}As_{15}H_{39}$ | 3.146 | 3.146 |
| 5 (Trough) | 2 × 1 | Al + $Ga_5As_6H_{16}$ | 3.115 | 3.115 |
| 5a | 4 × 2 | Al + $Ga_{19}As_{15}H_{39}$ | 2.890 | 2.890 |
| 5b | 4 × 2 | Al + $Ga_{19}As_{15}H_{39}$ | 2.890 | 2.890 |
| 6 (Cage) | 2 × 1 | Al + $Ga_4As_4H_{12}$ | 2.430 | 2.430 |
| | 2 × 1 | Al + $Ga_7As_6H_{19}$ | 2.430 | 2.430 |

Table 4. HOMO-LUMO gap (in eV) vs. cluster size and smmetry with 6-311++G** basis on Al.

| Sites | Symmetry | Cluster | Gap | Cluster | Gap | ΔGap |
|---|---|---|---|---|---|---|
| 1 (Top) | | | | | | |
| 1a | 2 × 1 | $Ga_4As_4H_{12}$ | 7.462 | Al + $Ga_4As_4H_{12}$ | 5.638 | -1.824 |
| 1b | 2 × 1 | $Ga_4As_4H_{12}$ | 7.462 | Al + $Ga_4As_4H_{12}$ | 5.644 | -1.818 |
| 1a | 2 × 1 | $Ga_7As_6H_{19}$ | 2.057 | Al + $Ga_7As_6H_{19}$ | 5.735 | 3.678 |
| 1b | 2 × 1 | $Ga_7As_6H_{19}$ | 2.057 | Al + $Ga_7As_6H_{19}$ | 5.468 | 3.411 |
| 1b | 4 × 2 | $Ga_{19}As_{15}H_{39}$ | 2.385 | Al + $Ga_{19}As_{15}H_{39}$ | 3.766 | 1.381 |
| 2 (Bridge) | 2 × 1 | $Ga_4As_4H_{12}$ | 7.462 | Al + $Ga_4As_4H_{12}$ | 5.504 | -1.958 |
| | 2 × 1 | $Ga_7As_6H_{19}$ | 2.057 | Al + $Ga_7As_6H_{19}$ | 6.060 | 4.003 |
| | 4 × 2 | $Ga_{19}As_{15}H_{39}$ | 2.385 | Al + $Ga_{19}As_{15}H_{39}$ | 2.659 | 0.274 |
| 3 (Hollow) | 2 × 1 | $Ga_7As_6H_{19}$ | 2.057 | Al + $Ga_7As_6H_{19}$ | 6.032 | 3.975 |
| | 4 × 2 | $Ga_{19}As_{15}H_{39}$ | 2.385 | Al + $Ga_{19}As_{15}H_{39}$ | 3.602 | 1.217 |
| 4 (Cave) | 2 × 1 | $Ga_7As_6H_{16}$ | 3.698 | Al + $Ga_7As_6H_{16}$ | 4.400 | 0.702 |
| | 4 × 2 | $Ga_{19}As_{15}H_{39}$ | 2.385 | Al + $Ga_{19}As_{15}H_{39}$ | 2.530 | 0.145 |
| 5 (Trough) | 2 × 1 | $Ga_5As_6H_{16}$ | 4.584 | Al + $Ga_5As_6H_{16}$ | 5.272 | 0.688 |
| 5a | 4 × 2 | $Ga_{19}As_{15}H_{39}$ | 2.385 | Al + $Ga_{19}As_{15}H_{39}$ | 2.693 | 0.308 |
| 5b | 4 × 2 | $Ga_{19}As_{15}H_{39}$ | 2.385 | Al + $Ga_{19}As_{15}H_{39}$ | 2.684 | 0.229 |
| 6 (Cage) | 2 × 1 | $Ga_4As_4H_{12}$ | 7.462 | Al + $Ga_4As_4H_{12}$ | 4.733 | -2.729 |
| | 2 × 1 | $Ga_7As_6H_{19}$ | 2.057 | Al + $Ga_7As_6H_{19}$ | 5.795 | 3.783 |

Table 5. HOMO-LUMO gap ( in eV) vs. cluster size and symmetry with H-W on Al adatom.

| Sites | Symmetry | Cluster | Gap | Cluster | Gap | ΔGap |
|---|---|---|---|---|---|---|
| 1 (Top) | | | | | | |
| 1a | 2 × 1 | $Ga_4As_4H_{12}$ | 7.462 | Al + $Ga_4As_4H_{12}$ | 6.334 | -1.128 |
| 1b | 2 × 1 | $Ga_4As_4H_{12}$ | 7.462 | Al + $Ga_4As_4H_{12}$ | 6.334 | -1.128 |
| 1a | 2 × 1 | $Ga_7As_6H_{19}$ | 2.057 | Al + $Ga_7As_6H_{19}$ | 5.726 | 3.669 |
| 1b | 2 × 1 | $Ga_7As_6H_{19}$ | 2.057 | Al + $Ga_7As_6H_{19}$ | 5.734 | 3.677 |
| 1b | 4 × 2 | $Ga_{19}As_{15}H_{39}$ | 2.385 | Al + $Ga_{19}As_{15}H_{39}$ | 3.218 | 0.833 |
| 2 (Bridge) | 2 × 1 | $Ga_4As_4H_{12}$ | 7.462 | Al + $Ga_4As_4H_{12}$ | 5.674 | -1.788 |
| | 2 × 1 | $Ga_7As_6H_{19}$ | 2.057 | Al + $Ga_7As_6H_{19}$ | 6.416 | 4.359 |
| | 4 × 2 | $Ga_{19}As_{15}H_{39}$ | 2.385 | Al + $Ga_{19}As_{15}H_{39}$ | 3.220 | 0.835 |
| 3 (Hollow) | 2 × 1 | $Ga_7As_6H_{19}$ | 2.057 | Al + $Ga_7As_6H_{19}$ | 6.145 | 4.088 |
| | 4 × 2 | $Ga_{19}As_{15}H_{39}$ | 2.385 | Al + $Ga_{19}As_{15}H_{39}$ | 3.407 | 1.022 |
| 4 (Cave) | 2 × 1 | $Ga_7As_6H_{16}$ | 3.698 | Al + $Ga_7As_6H_{16}$ | 4.410 | 0.712 |
| | 4 × 2 | $Ga_{19}As_{15}H_{39}$ | 2.385 | Al + $Ga_{19}As_{15}H_{39}$ | 2.631 | 0.246 |
| 5 (Trough) | 2 × 1 | $Ga_5As_6H_{16}$ | 4.584 | Al + $Ga_5As_6H_{16}$ | 5.244 | 0.660 |
| 5a | 4 × 2 | $Ga_{19}As_{15}H_{39}$ | 2.385 | Al + $Ga_{19}As_{15}H_{39}$ | 2.692 | 0.307 |
| 5b | 4 × 2 | $Ga_{19}As_{15}H_{39}$ | 2.385 | Al + $Ga_{19}As_{15}H_{39}$ | 2.684 | 0.299 |
| 6 (Cage) | 2 × 1 | $Ga_4As_4H_{12}$ | 7.462 | Al + $Ga_4As_4H_{12}$ | 4.925 | -2.537 |
| | 2 × 1 | $Ga_7As_6H_{19}$ | 2.057 | Al + $Ga_7As_6H_{19}$ | 6.220 | 4.163 |

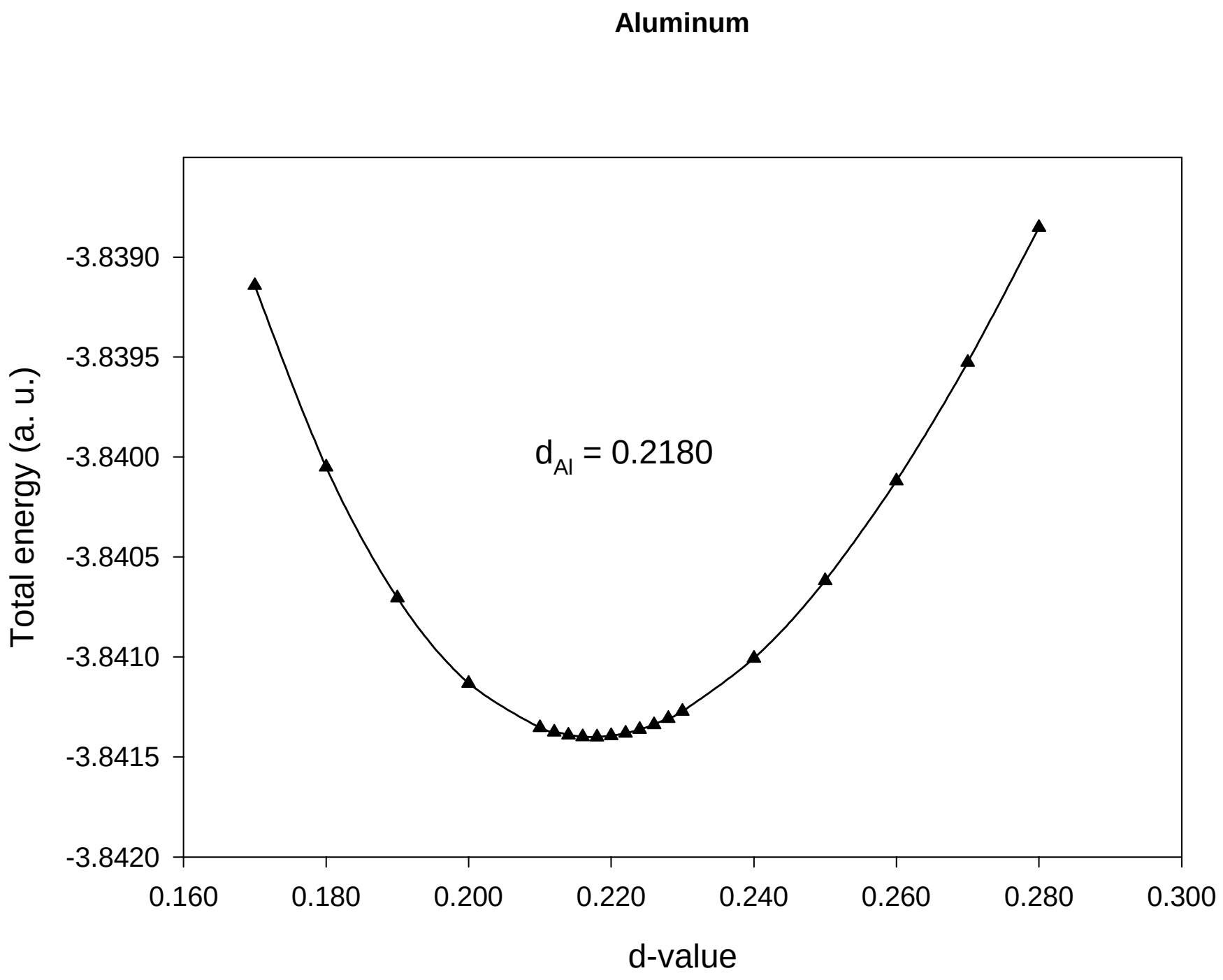

Fig. 1. Total energy (a.u.) versus the d-function exponent for Al.

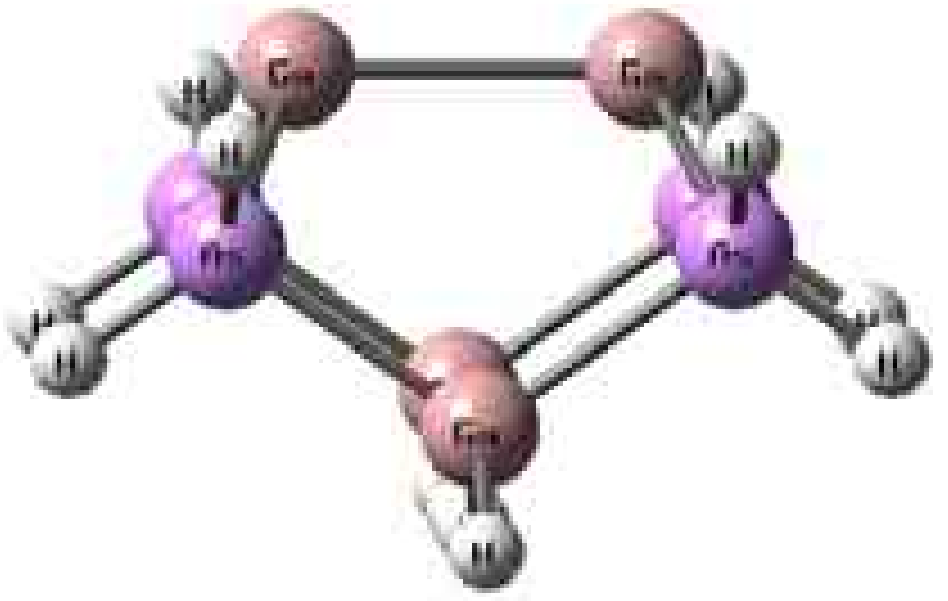

$Ga_4As_4H_{12}$ (2 × 1)

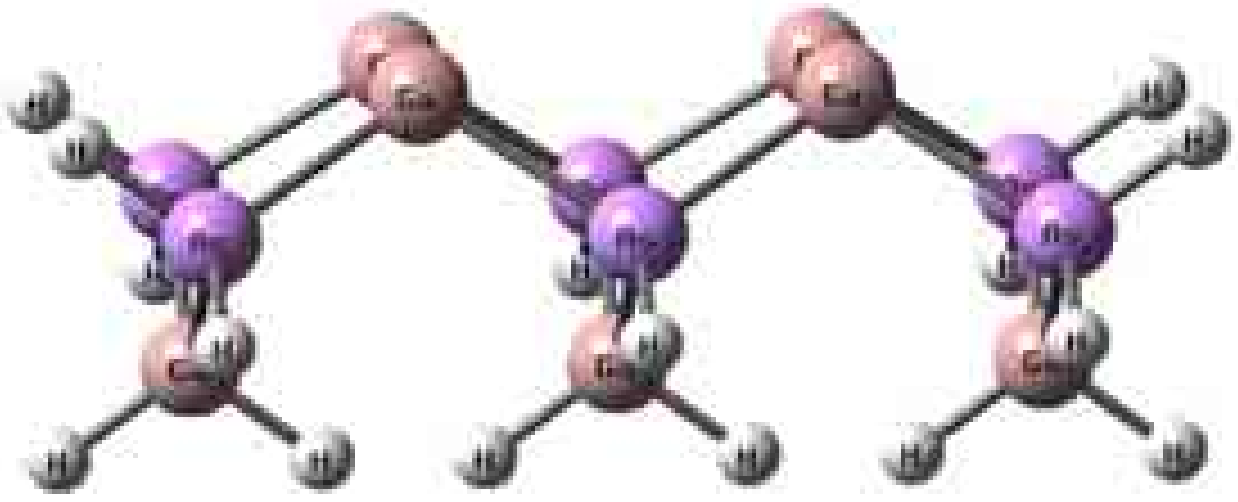

$Ga_5As_6H_{16}$ (2 × 1)

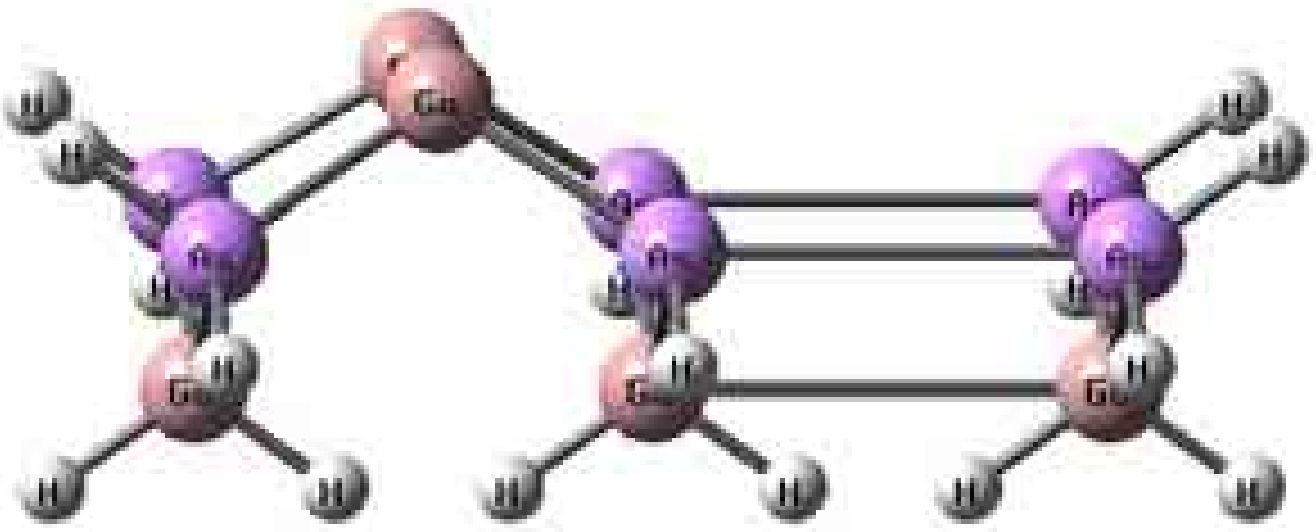

$Ga_7As_6H_{16}$ (2 × 1)

Fig. 2. GaAs(100) Clusters.

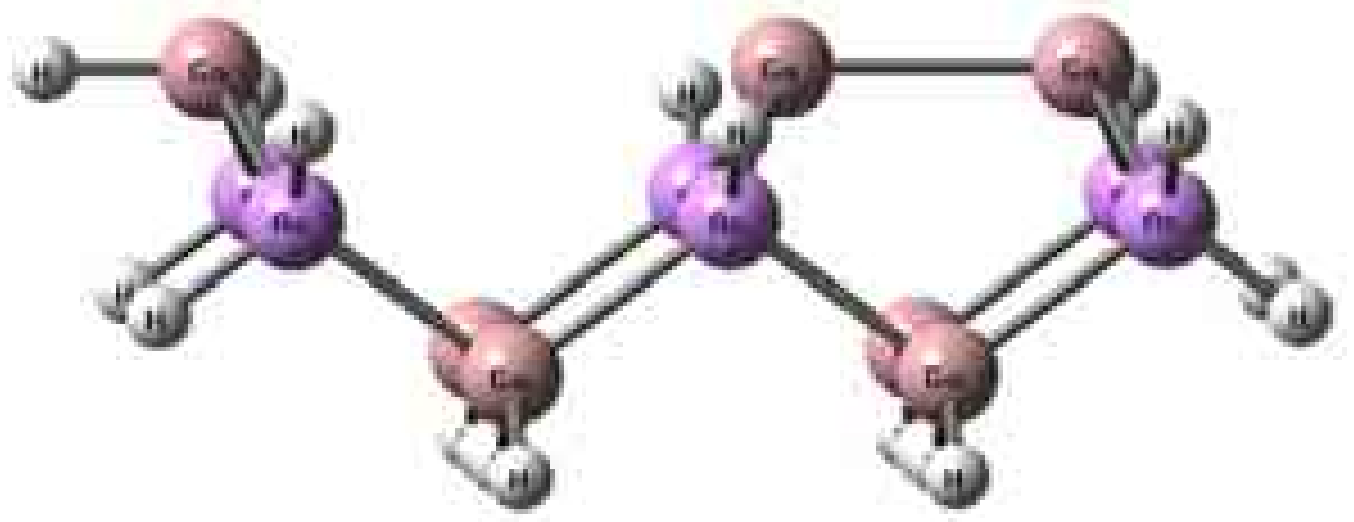

$Ga_7As_6H_{19}$ (2 × 1)

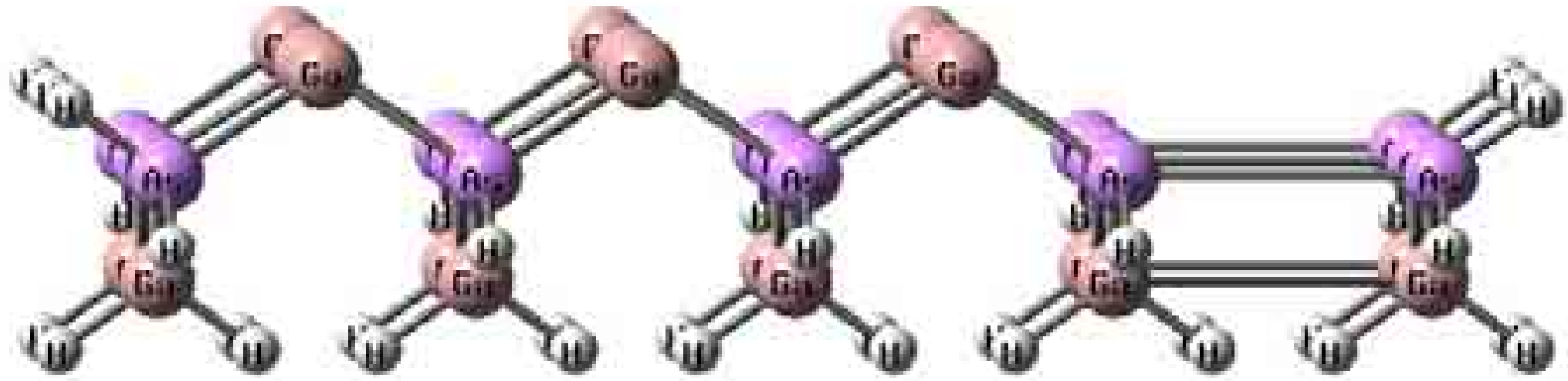

$Ga_{19}As_{15}H_{39}$ $\beta$(4 × 2)

Fig. 2 (cont.)

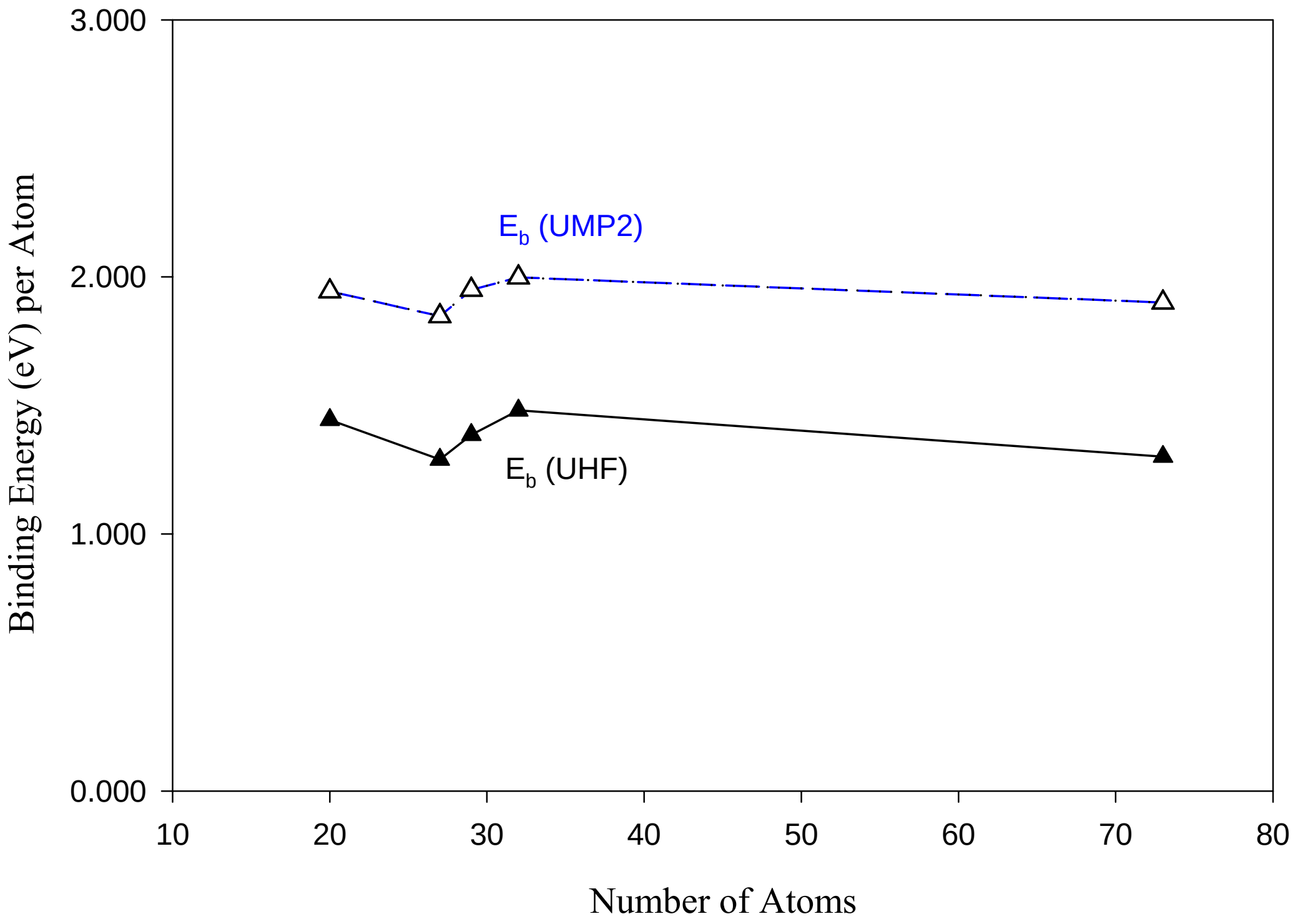

Fig. 3. Binding energy (eV) *vs.* number of atoms in the cluster.

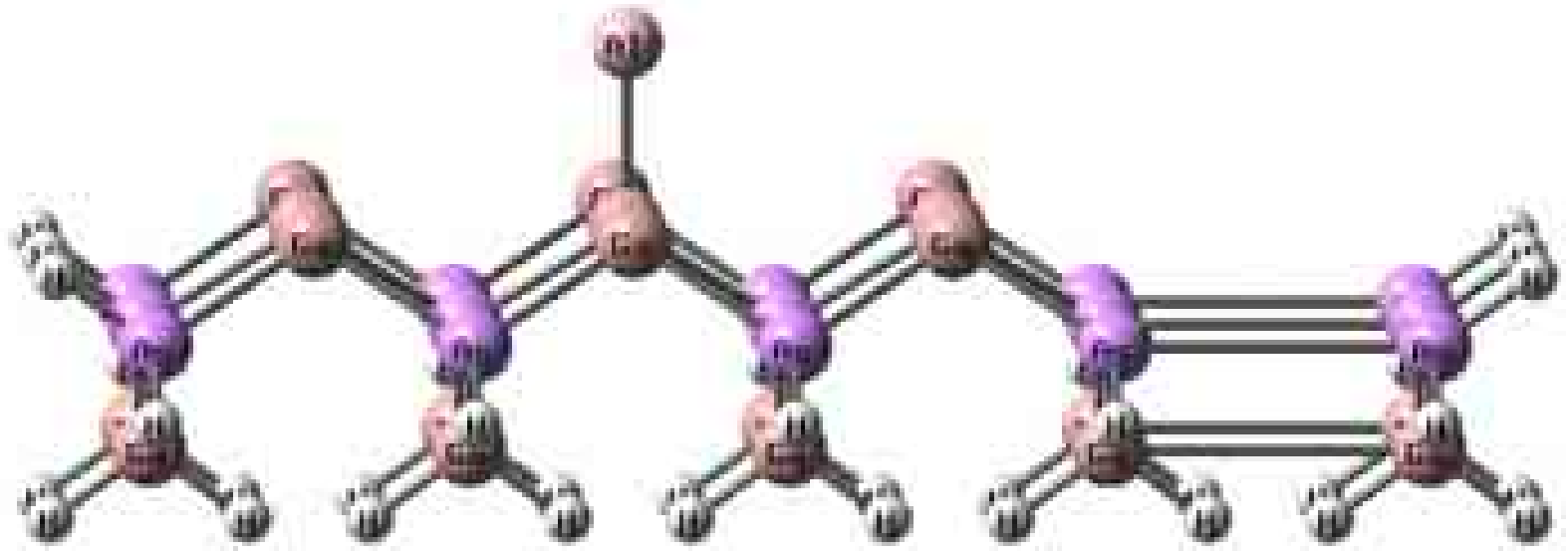

$Ga_{19}As_{15}H_{39}$ + Al Top Site 1b.

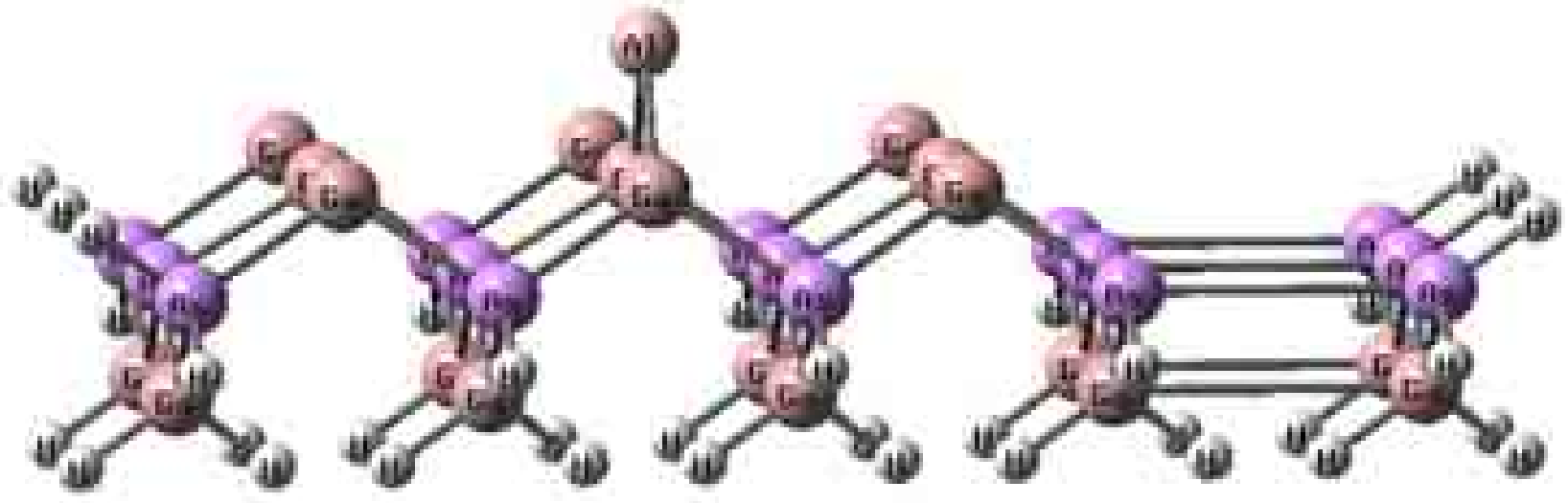

$Ga_{19}As_{15}H_{39}$ + Al Bridge Site 2.

Fig. 4. GaAs (100) chemisorbed clusters.

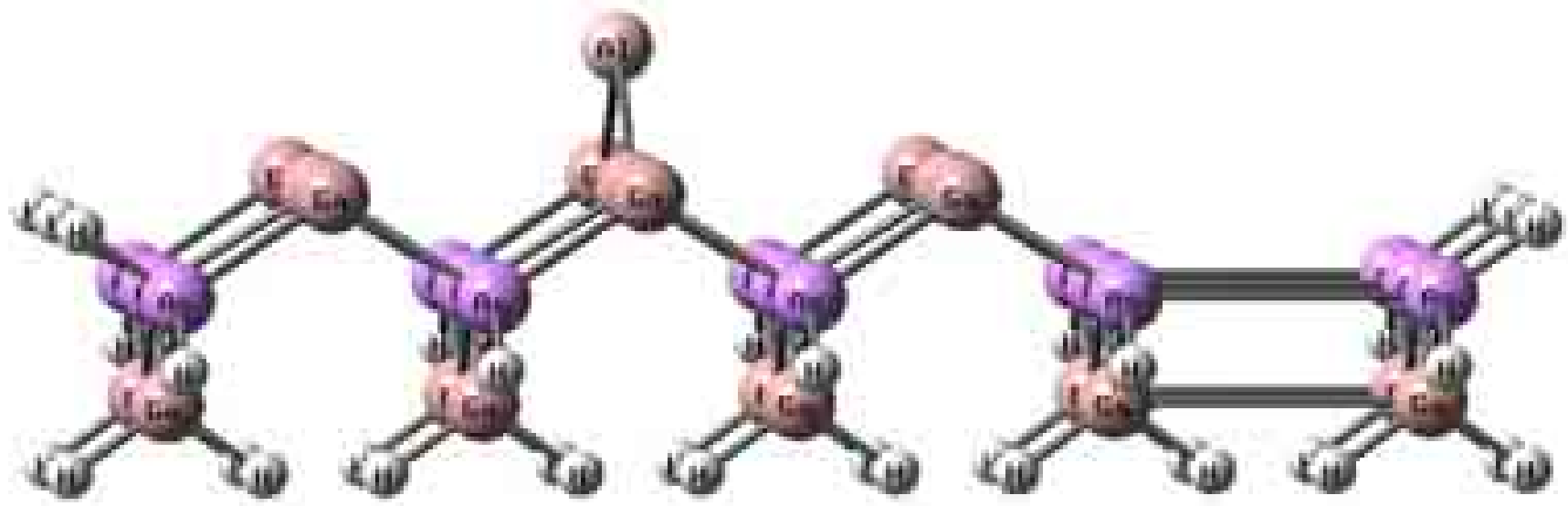

$Ga_{19}As_{15}H_{39}$ + Al Hollow Site 3.

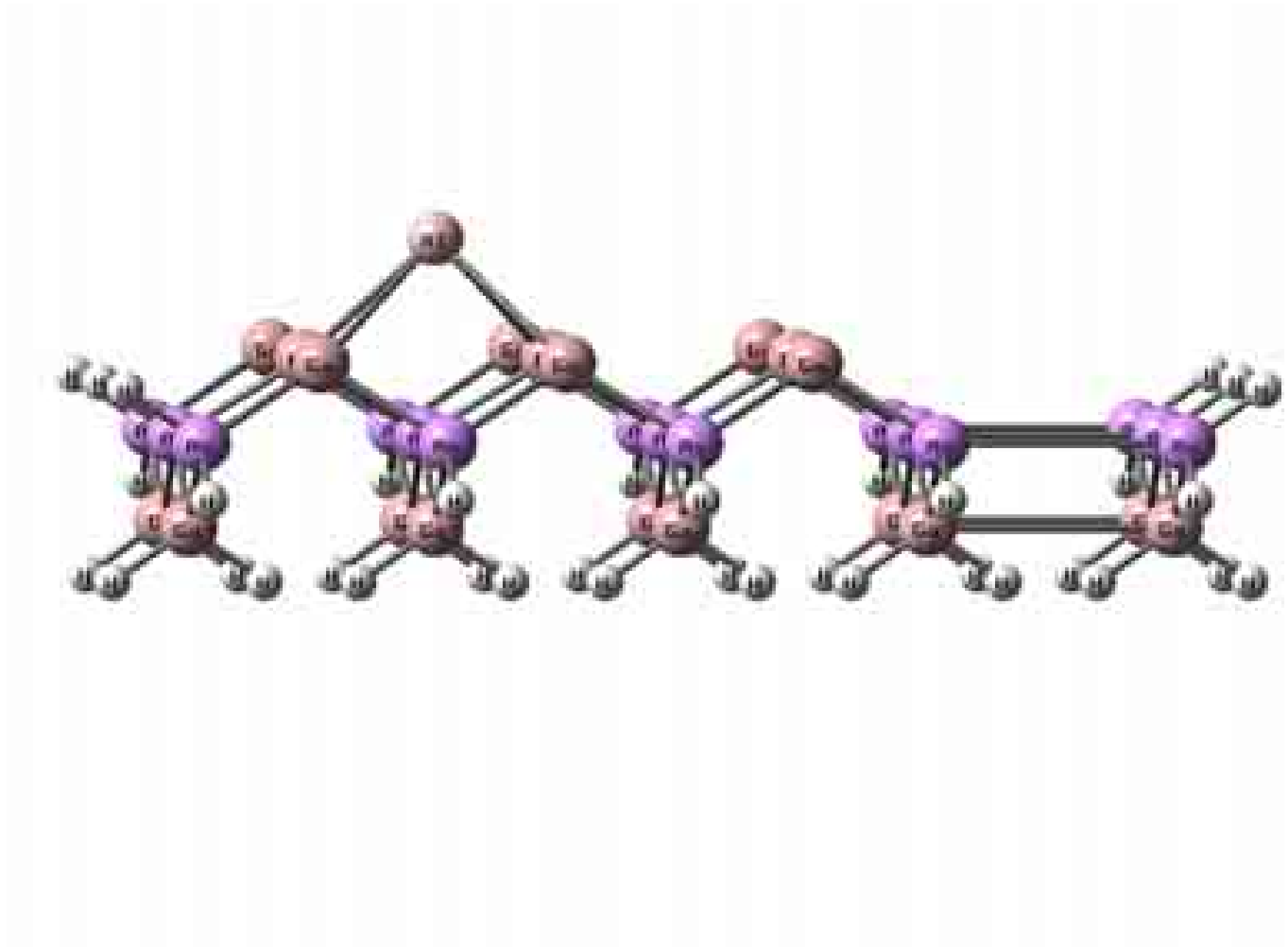

$Ga_{19}As_{15}H_{39}$ + Al Cave Site 4.

Fig. 4. (cont.)

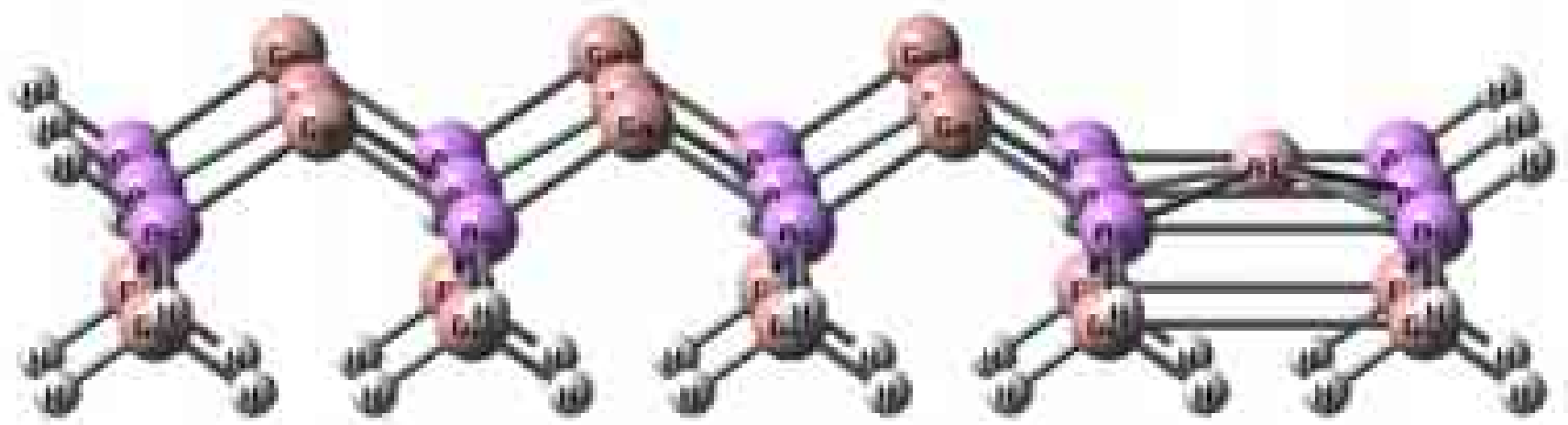

$Ga_{19}As_{15}H_{39}$ + Al Trough Site 5a.

Fig. 4. (cont.)

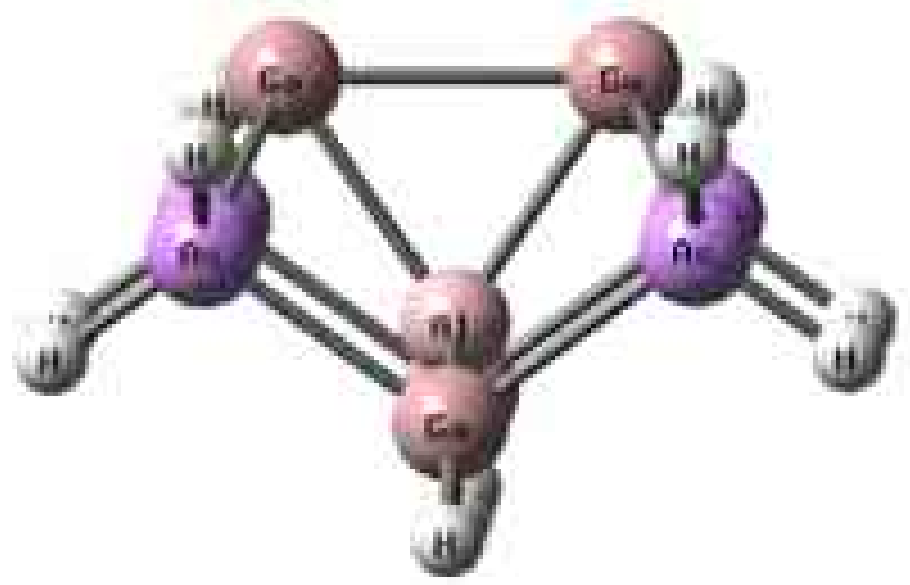

$Ga_4As_4H_{12}$ + Al Cage Site 6.

Fig. 5. GaAs (100) cage site.

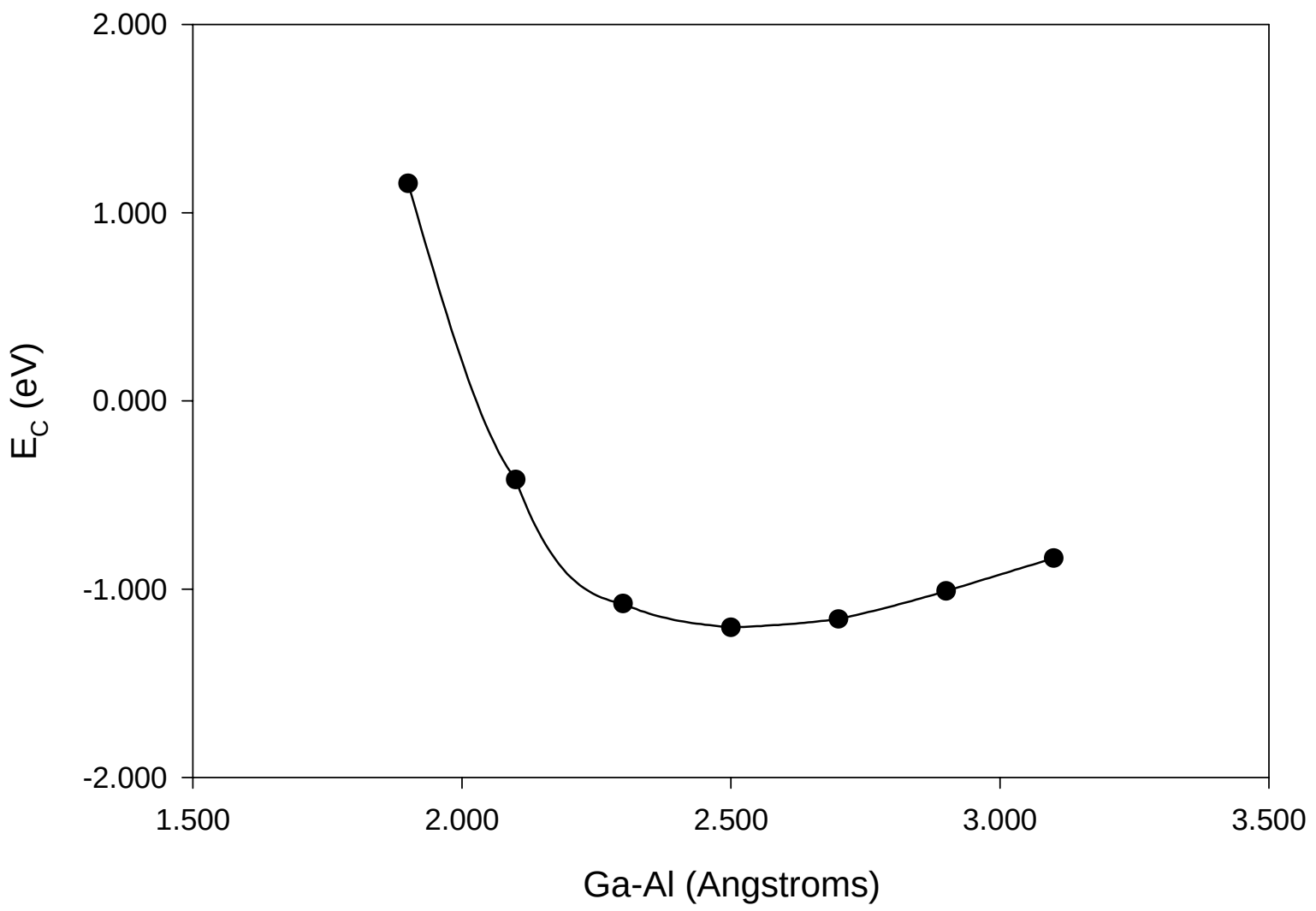

Fig. 6. Chemisorption energy vs. nearest surface neighbor bond length for $Ga_4As_4H_{12}$ + Al top Site 1a with 6-311++G** basis on Al.

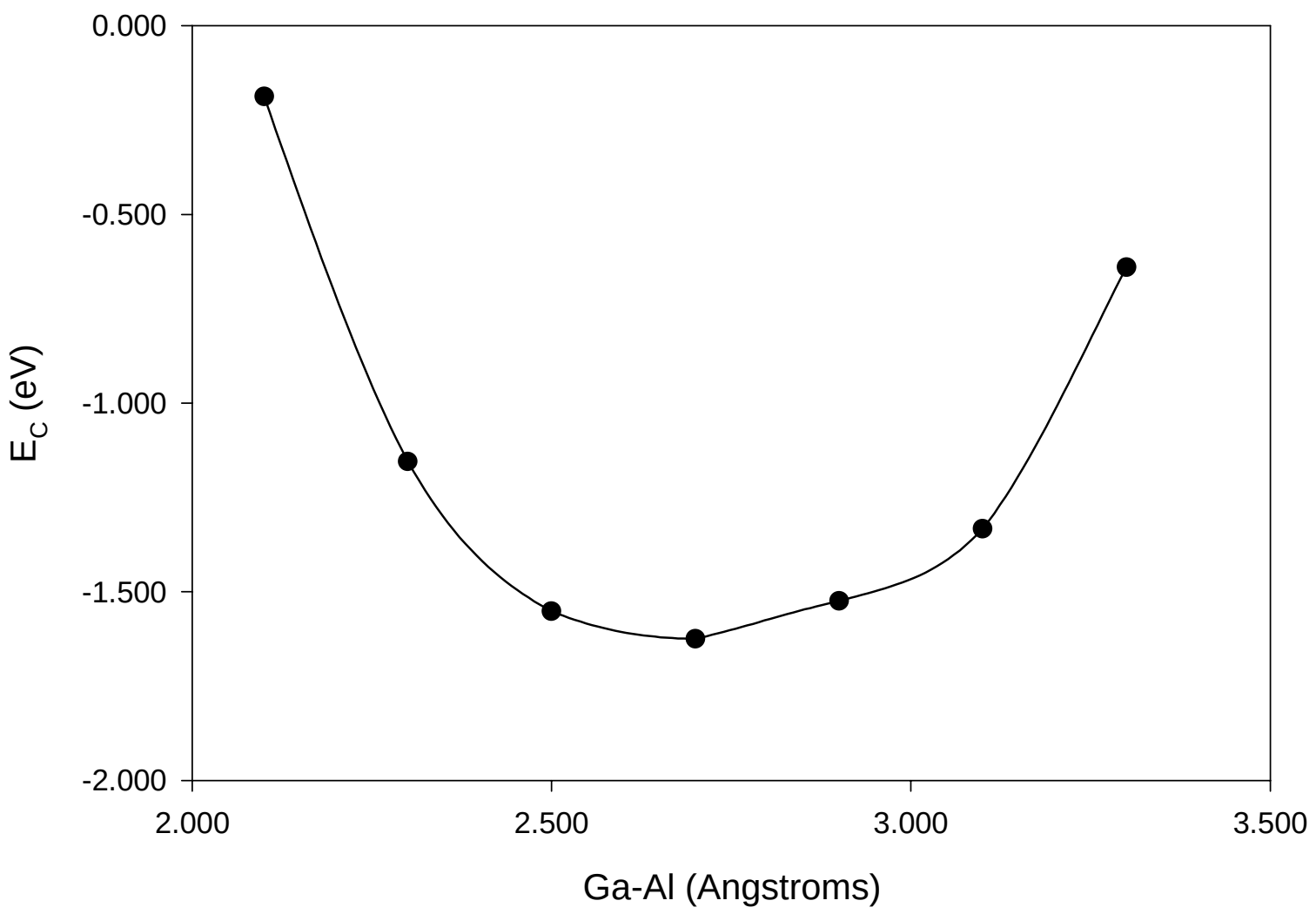

Fig. 7. Chemisorption energy vs. nearest surface neighbor bond length for $Ga_4As_4H_{12}$ + Al top site 1a with Hay-Wadt pseudopotential on Al.

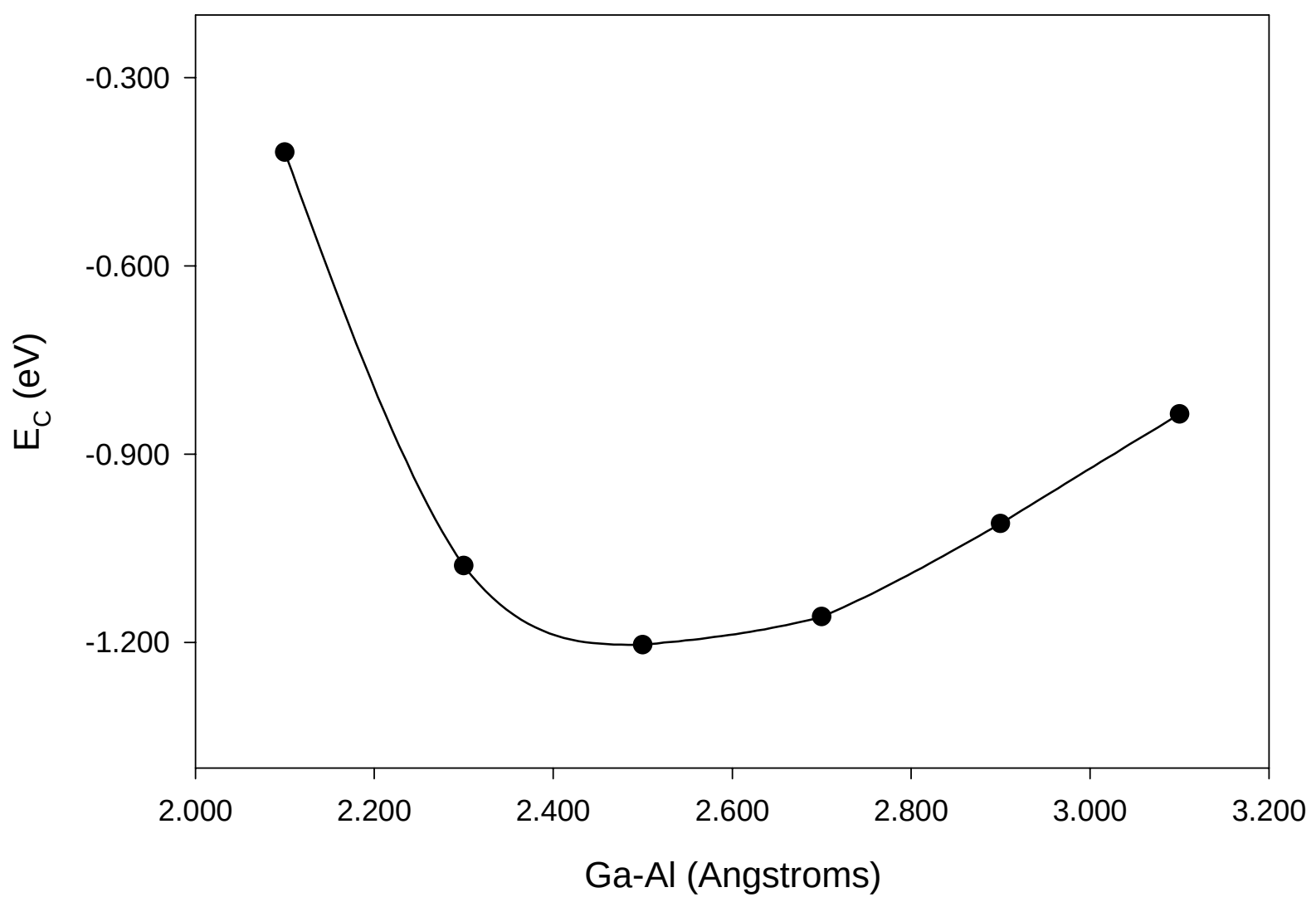

Fig. 8. Chemisorption energy vs. nearest surface neighbor bond length for $Ga_4As_4H_{12}$ + Al top site 1b with 6-311++G** basis on Al.

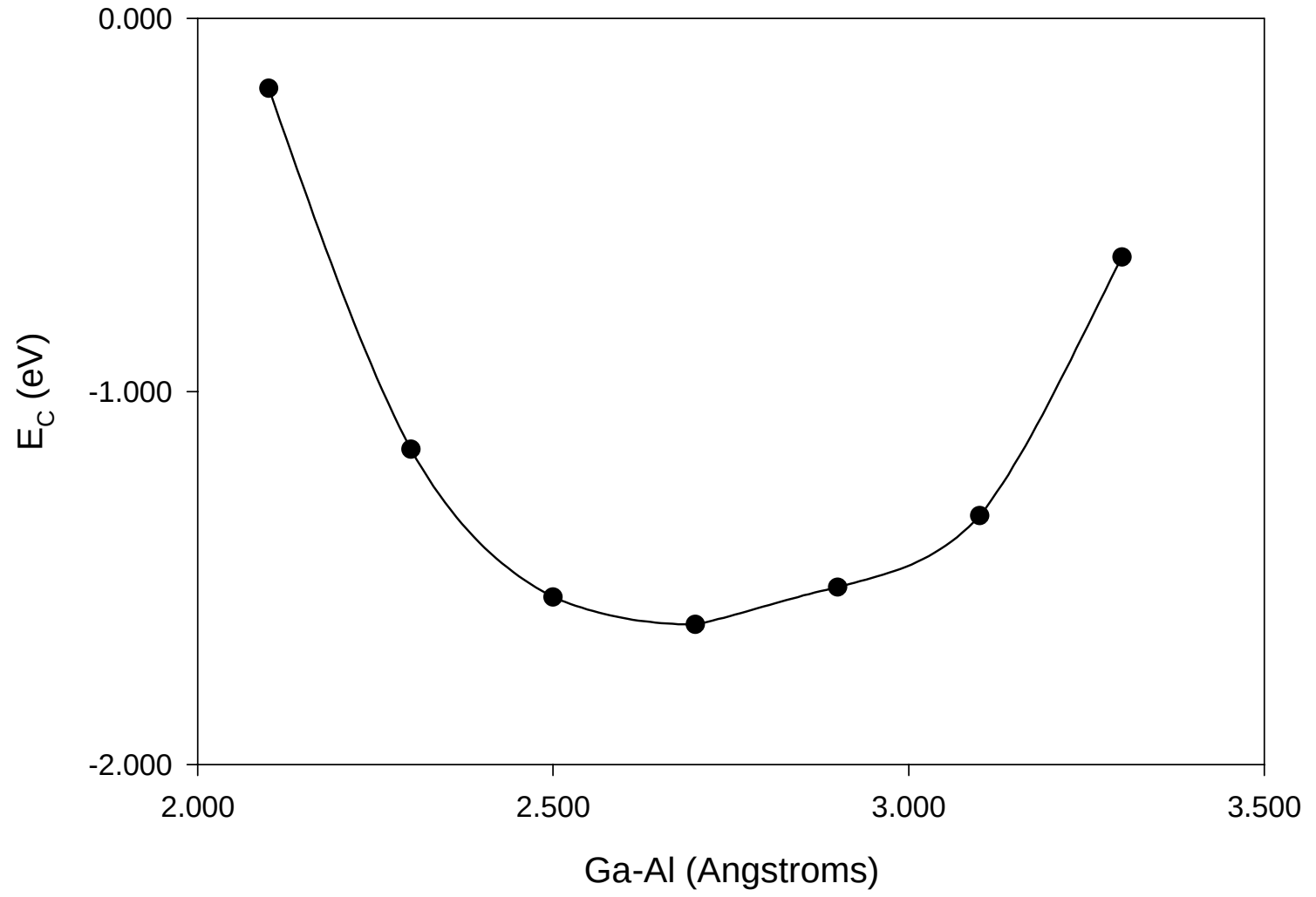

Fig. 9. Chemisorption energy vs. nearest surface neighbor bond length for $Ga_4As_4H_{12}$ + Al Top Site 1b with Hay-Wadt pseudopotential on Al.

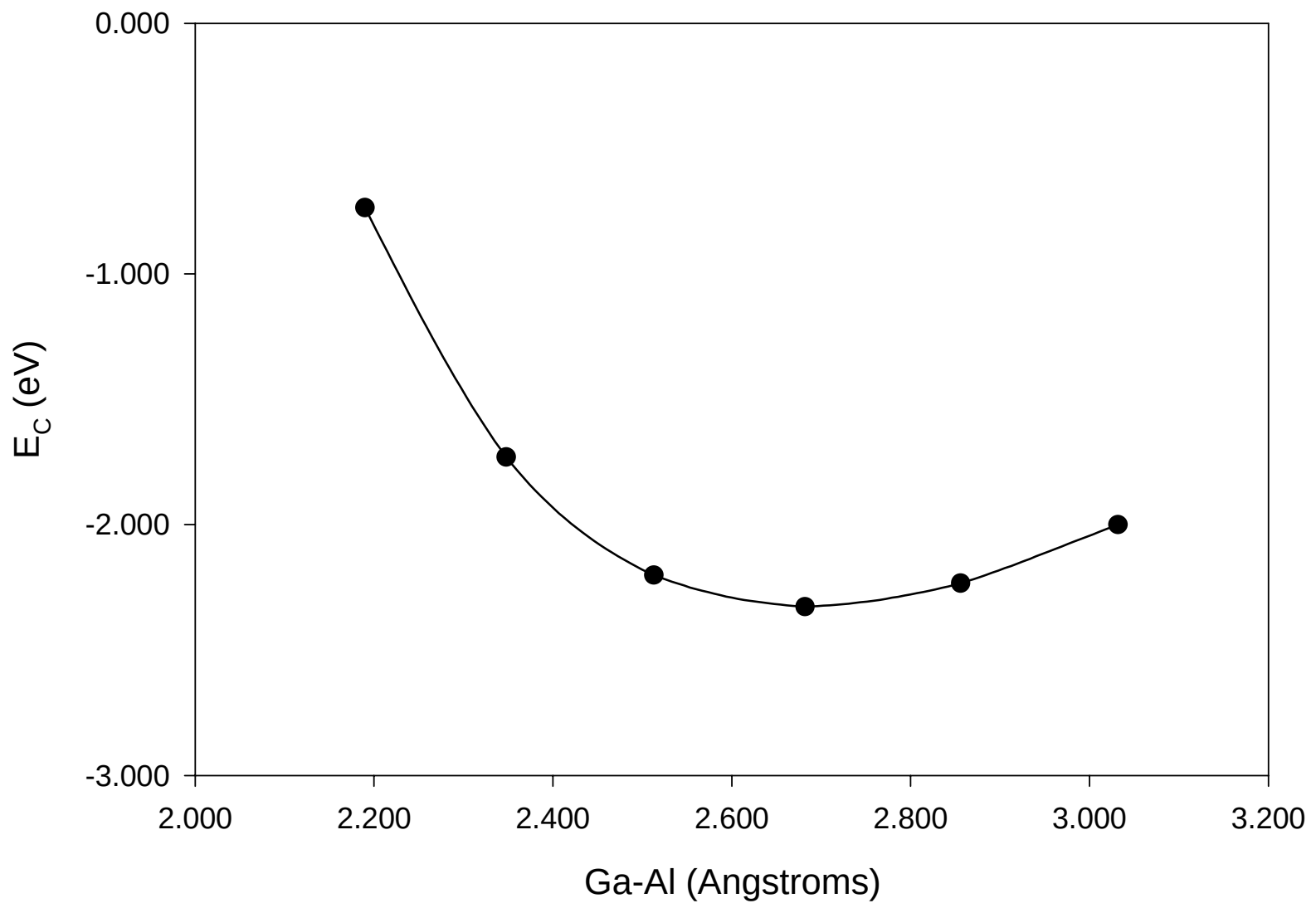

Fig. 10. Chemisorption energy vs. nearest surface neighbor bond length for $Ga_4As_4H_{12}$ + Al bridge site 2 with 6-311++G** basis on Al.

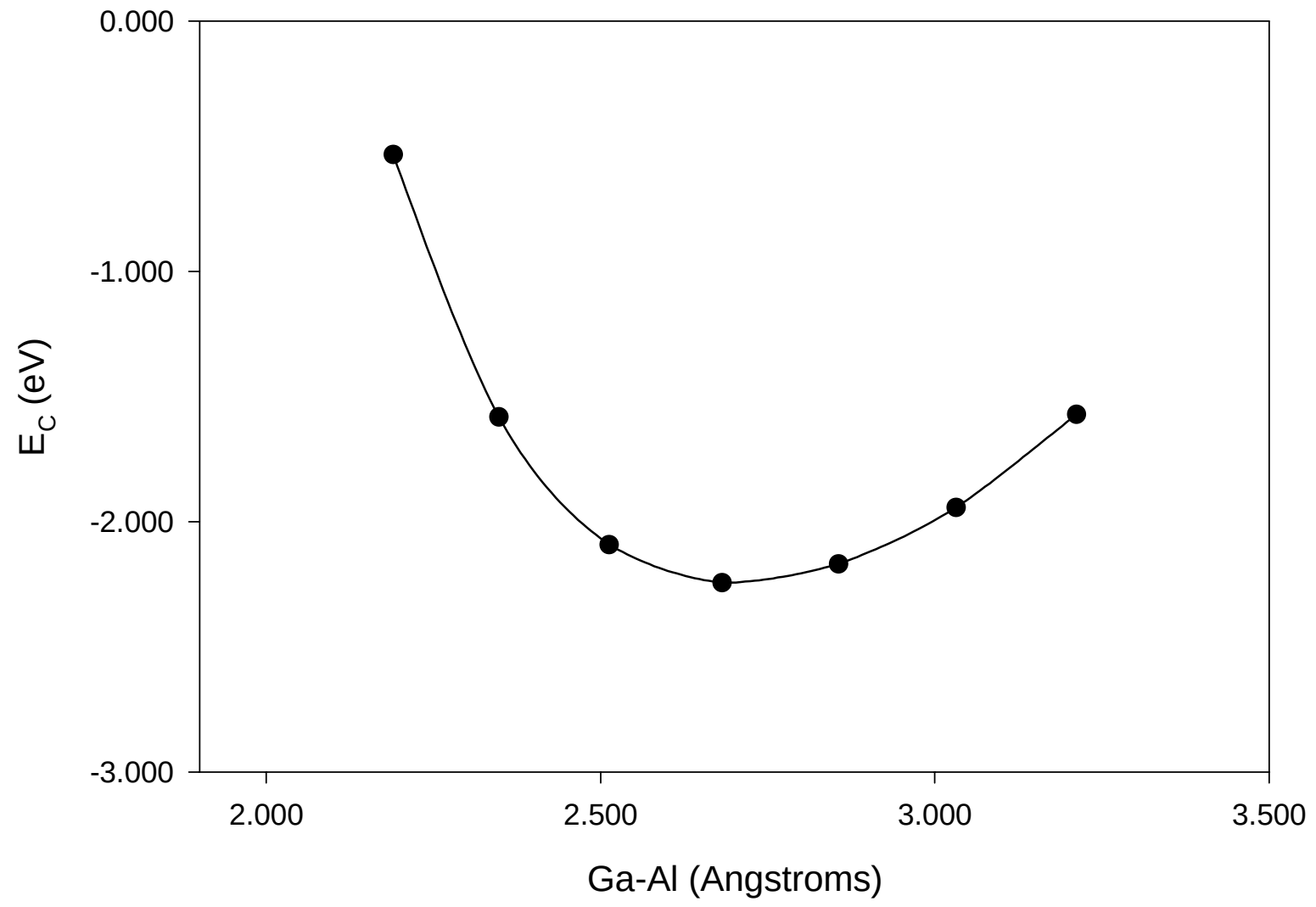

Fig. 11. Chemisorption energy vs. nearest surface neighbor bond length for $Ga_4As_4H_{12}$ + Al bridge site 2 with Hay-Wadt pseudopotential on Al.

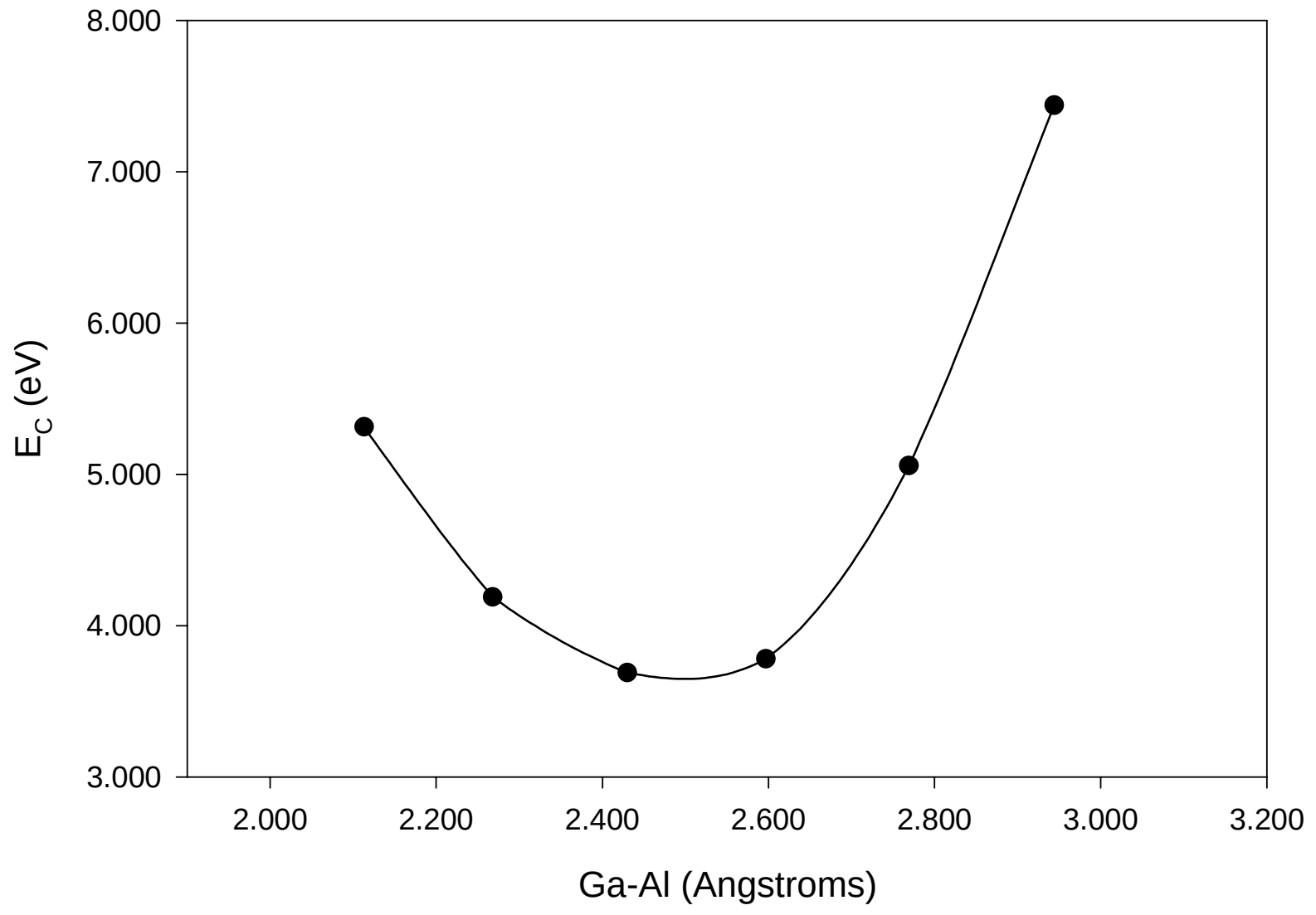

Fig.12. Chemisorption energy vs. nearest surface neighbor bond length for $Ga_4As_4H_{12}$ + Al cage site 6 with 6-311++G** basis on Al.

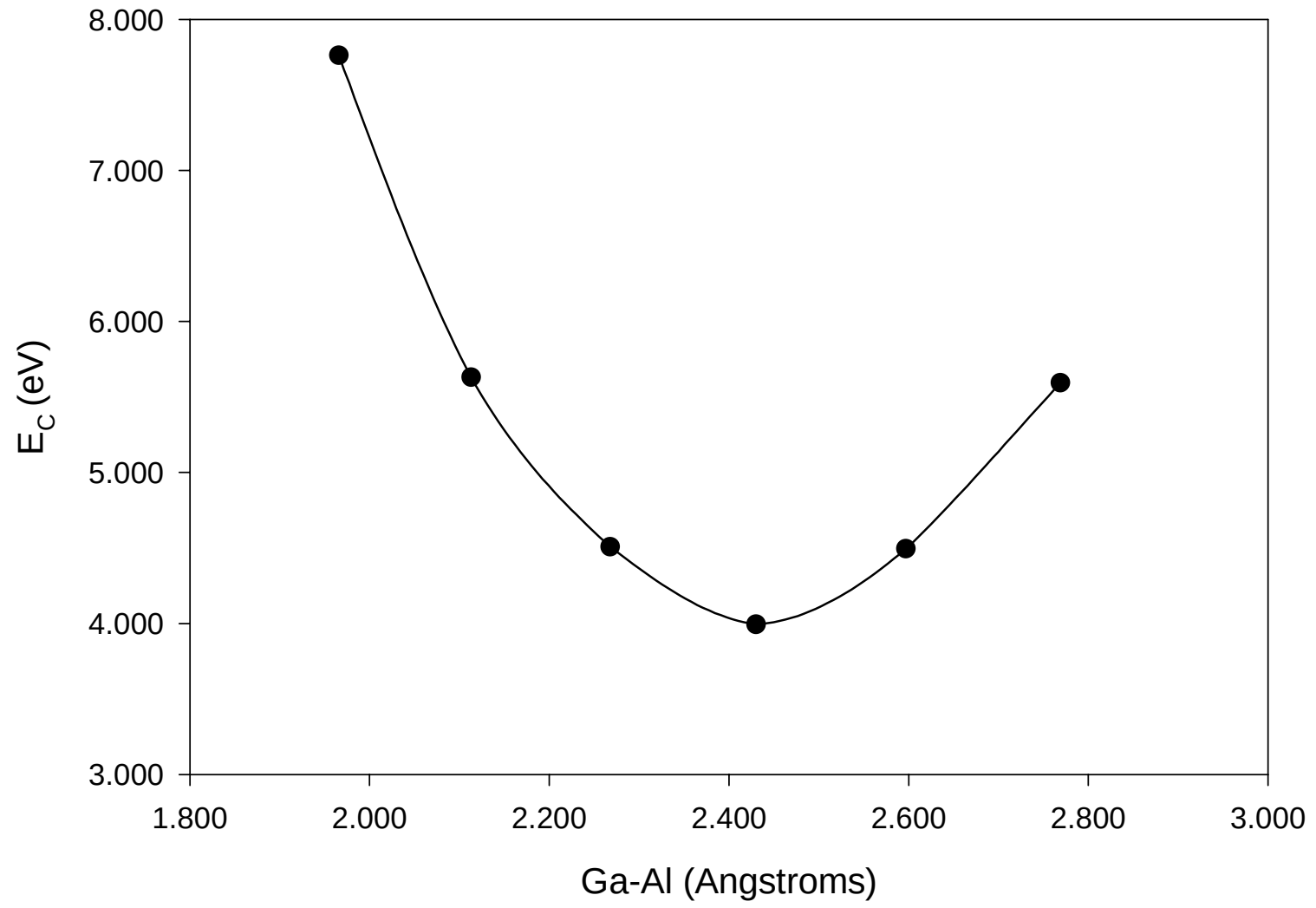

Fig. 13. Chemisorption energy vs. nearest surface neighbor bond length for $Ga_4As_4H_{12}$ + Al cage site 6 with Hay-Wadt pseudopotential on Al.

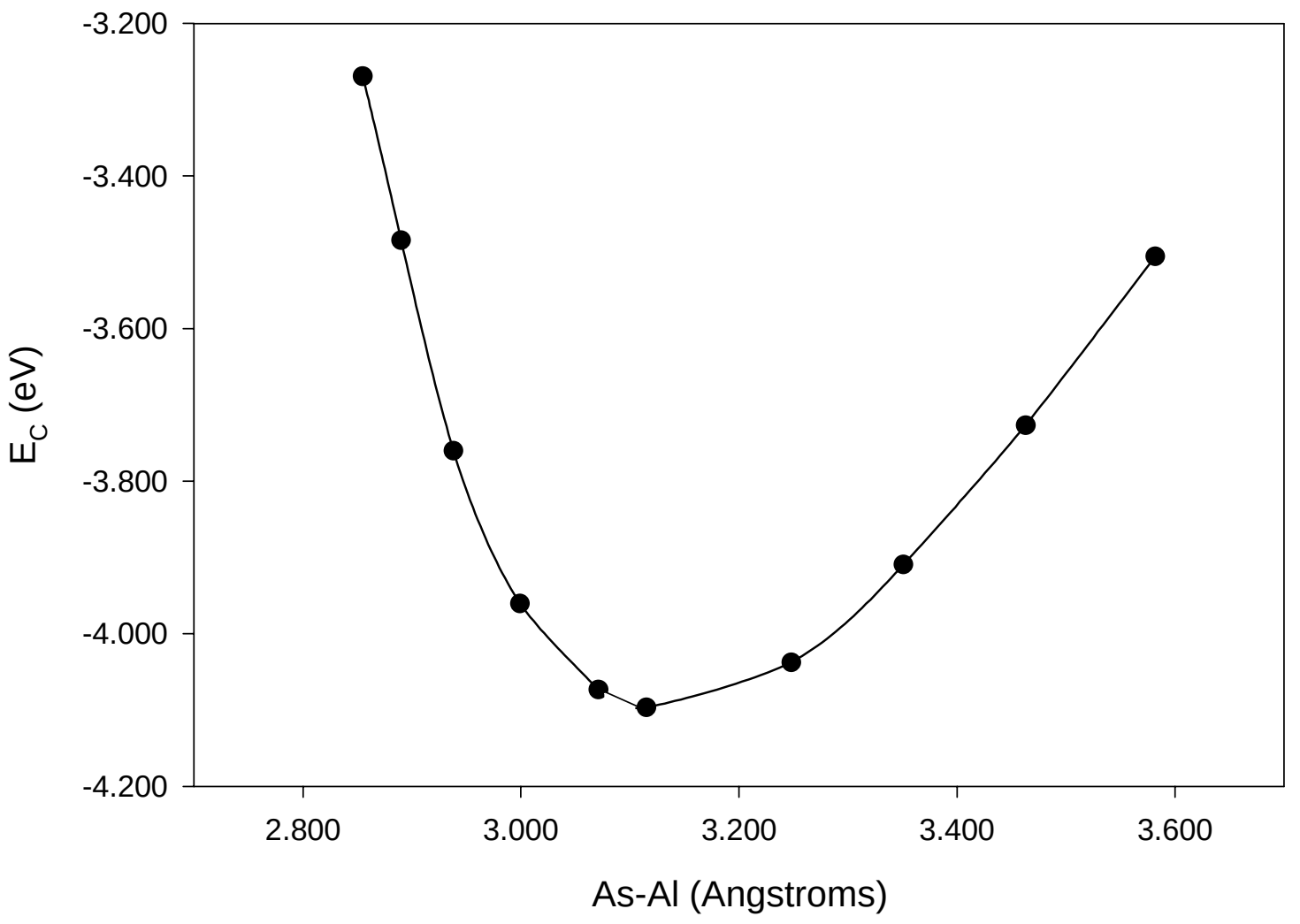

Fig. 14. Chemisorption energy vs. nearest surface neighbor bond length for $Ga_5As_6H_{16}$ + Al trough site 5 with 6-311++G** basis on Al.

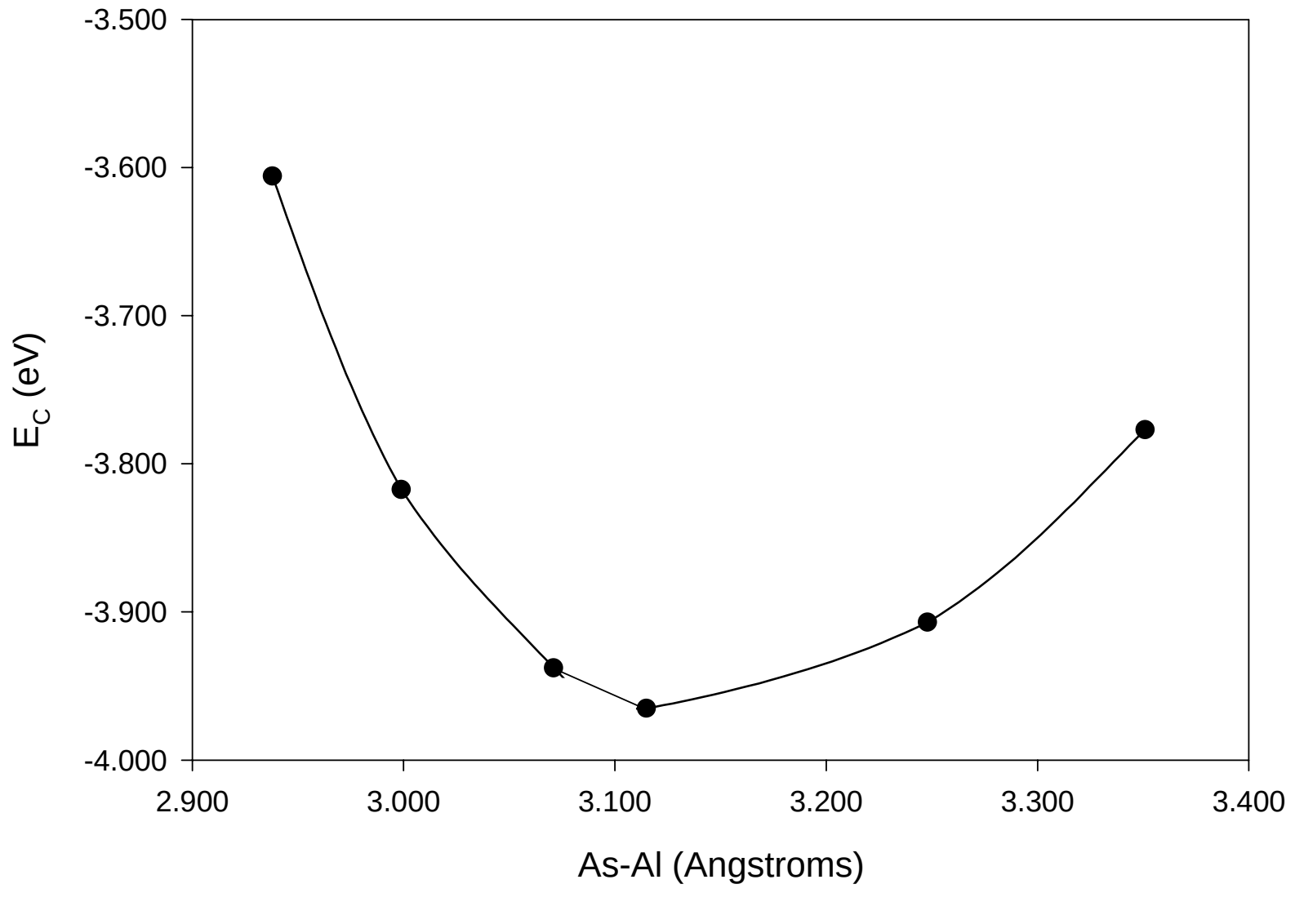

Fig. 15. Chemisorption energy vs. nearest surface neighbor bond length for $Ga_5As_6H_{16}$ + Al trough site 5 with Hay-Wadt pseudopotential on Al.

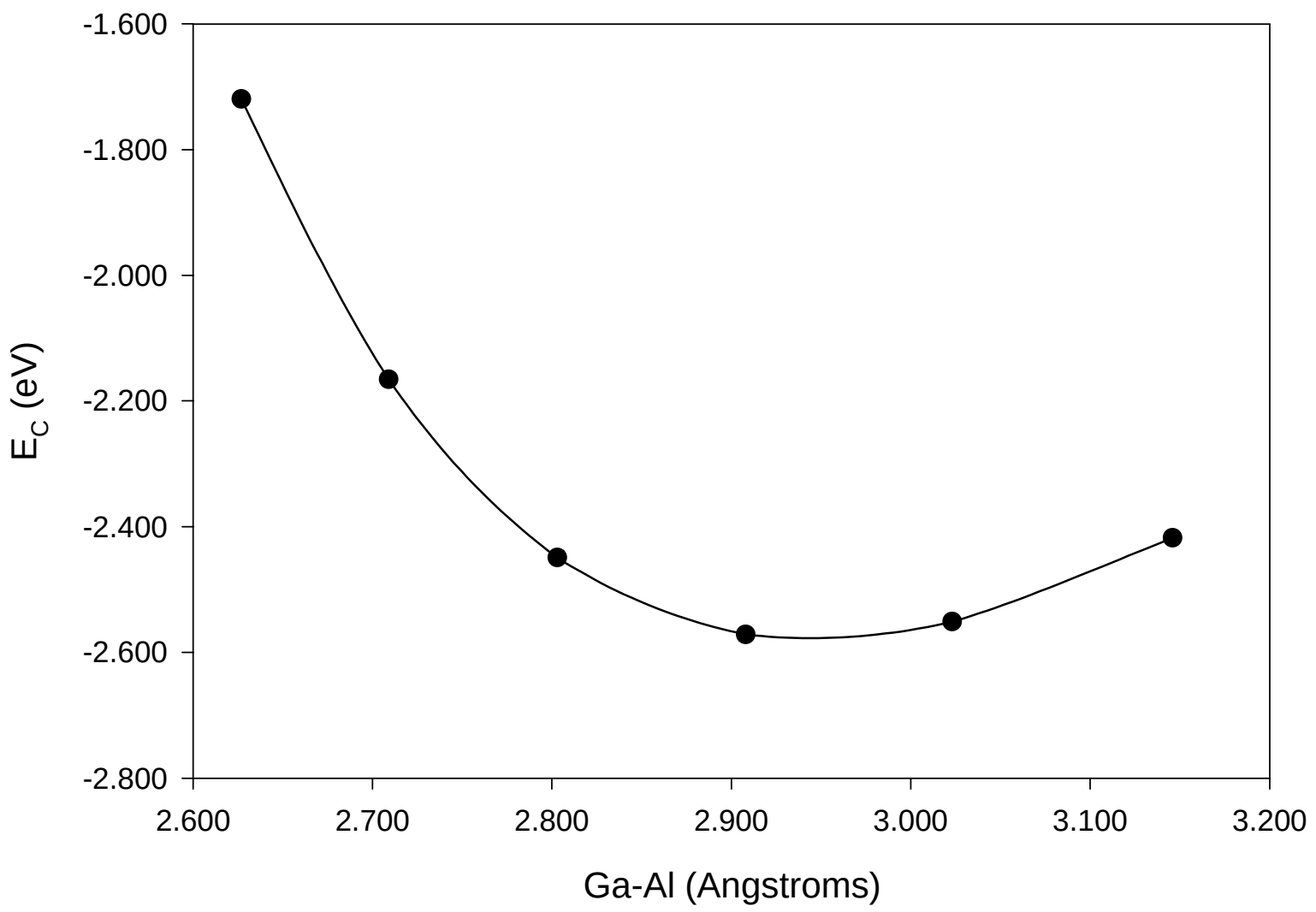

Fig. 16. Chemisorption energy vs. nearest surface neighbor bond length for $Ga_7As_6H_{16}$ + Al cave site 4 with 6-311++G** basis on Al.

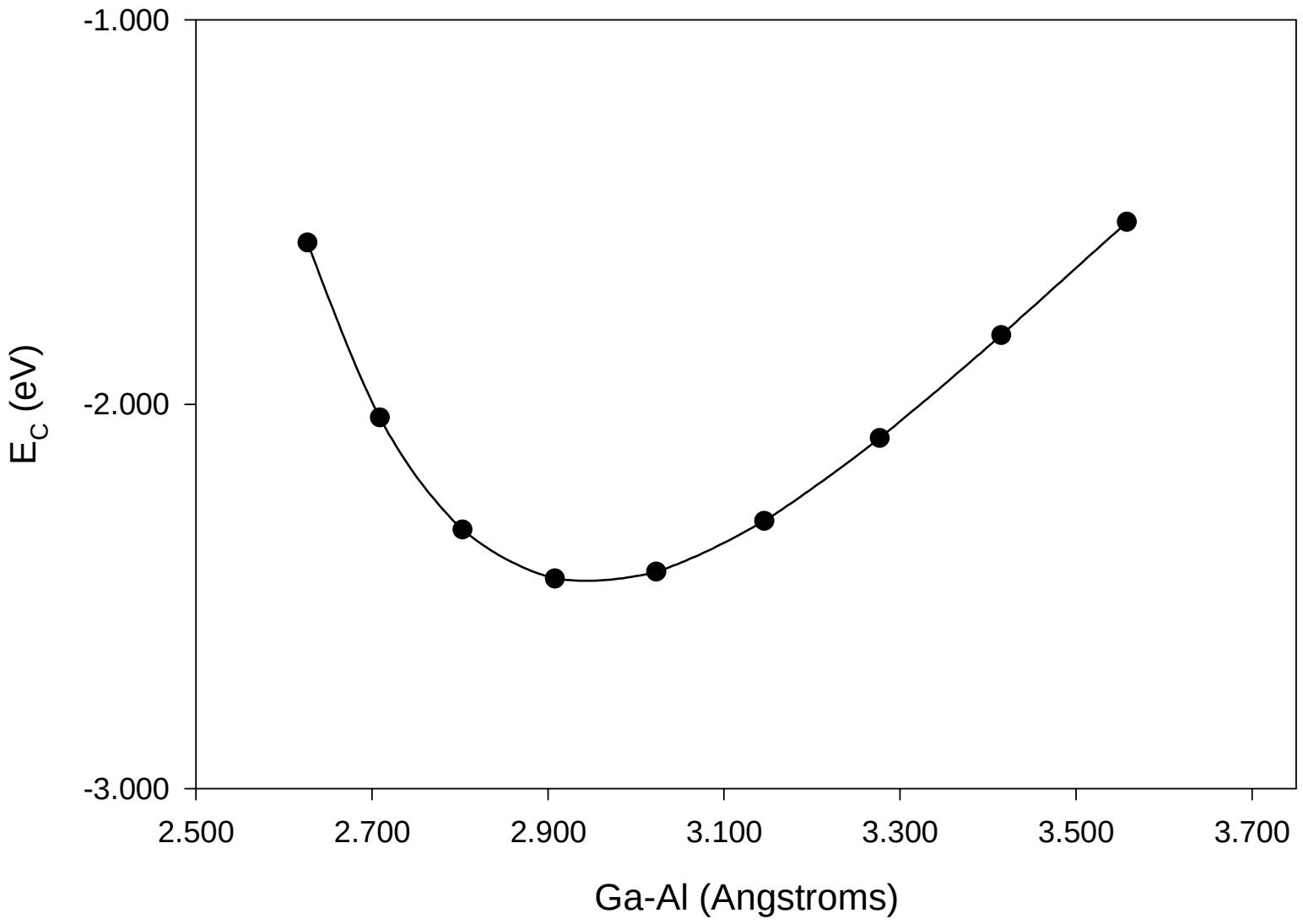

Fig. 17. Chemisorption Energy vs. Nearest Surface Neighbor Bond Length for $Ga_7As_6H_{16}$ + Al Cave Site 4 with Hay-Wadt Pseudopotential on Al.

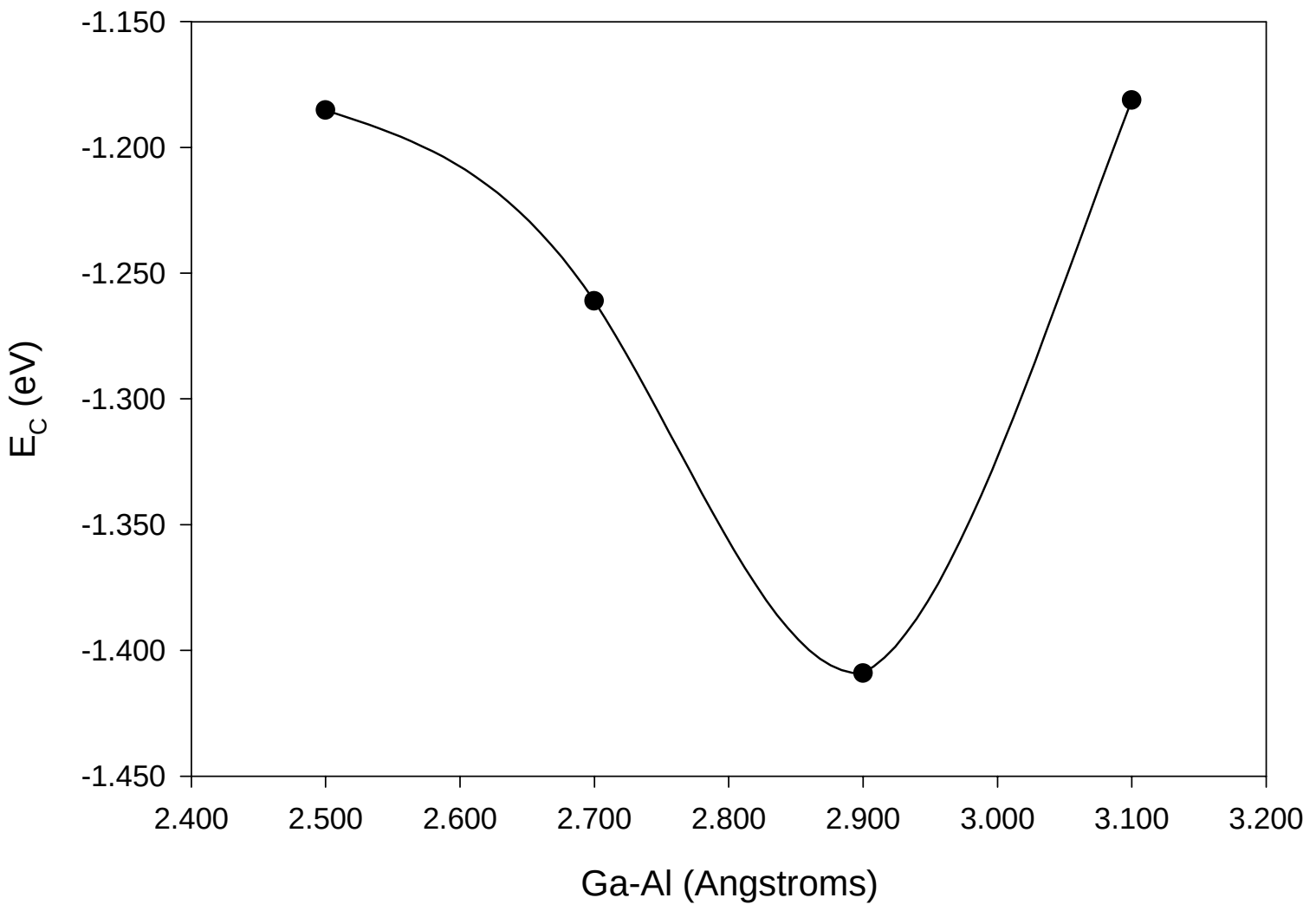

Fig. 18. Chemisorption energy vs. nearest surface neighbor bond length for $Ga_7As_6H_{19}$ + Al top site 1a with 6-311++G** basis on Al.

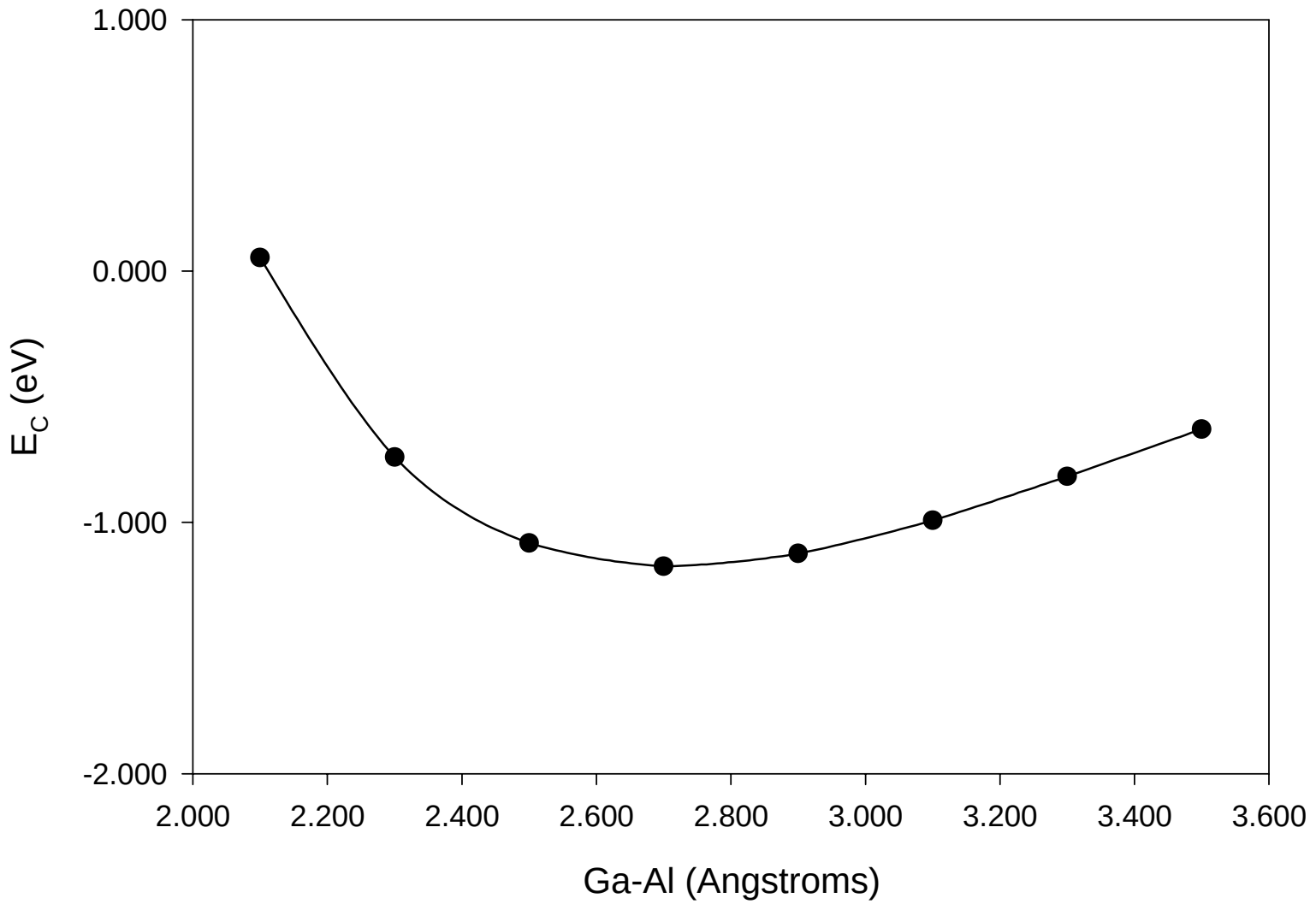

Fig. 19. Chemisorption energy vs. nearest surface neighbor bond length for $Ga_7As_6H_{19}$ + Al top site 1a with Hay-Wadt pseudopotential on Al.

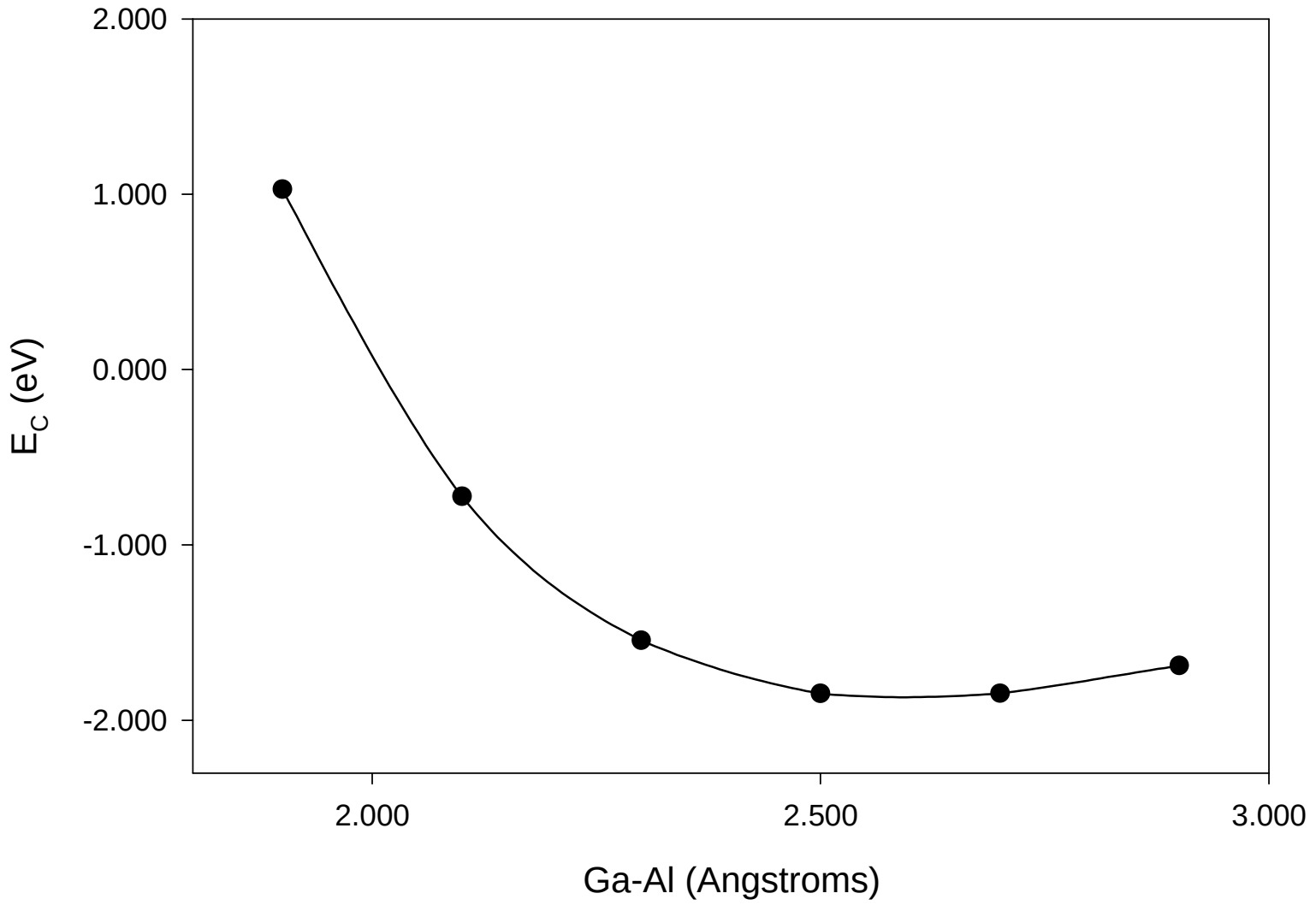

Fig. 20. Chemisorption energy vs. nearest surface neighbor bond length for $Ga_7As_6H_{19}$ + Al top site 1b with 6-311++G** basis on Al.

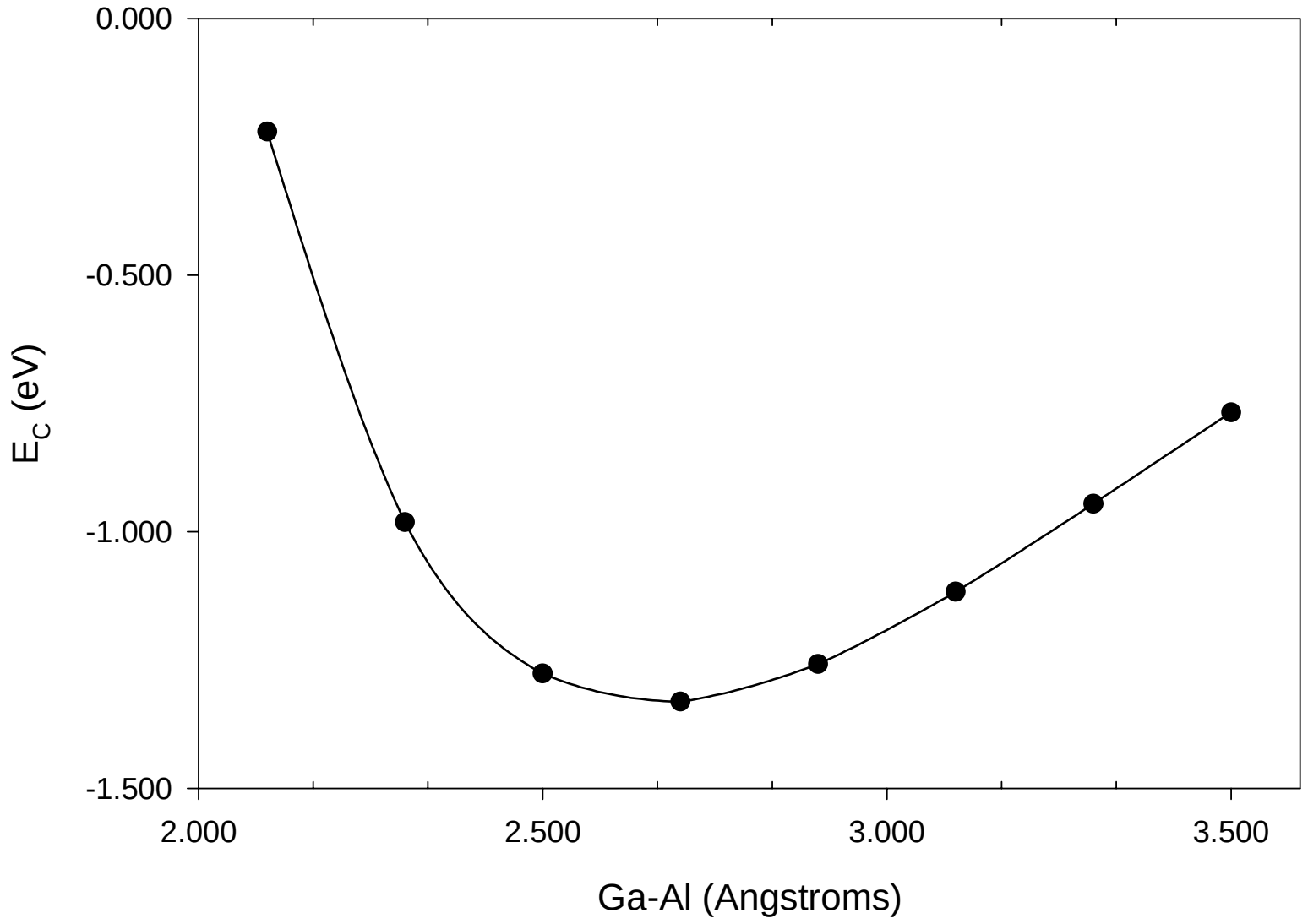

Fig. 21. Chemisorption energy vs. nearest surface neighbor bond length for $Ga_7As_6H_{19}$ + Al top site 1b with Hay-Wadt pseudopotential on Al.

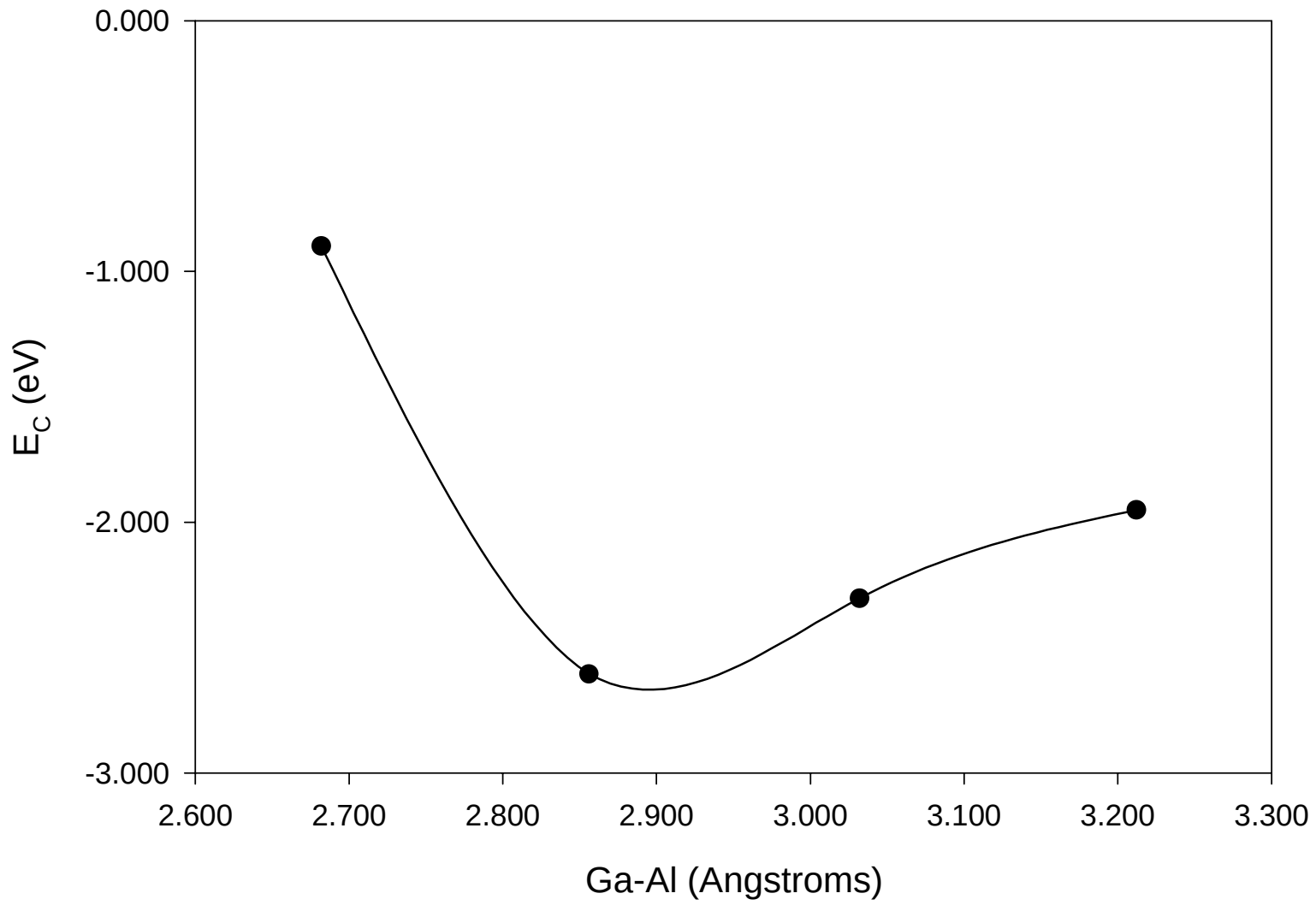

Fig. 22. Chemisorption energy vs. nearest surface neighbor bond length for $Ga_7As_6H_{19}$ + Al bridge site 2 with 6-311++G** basis on Al.

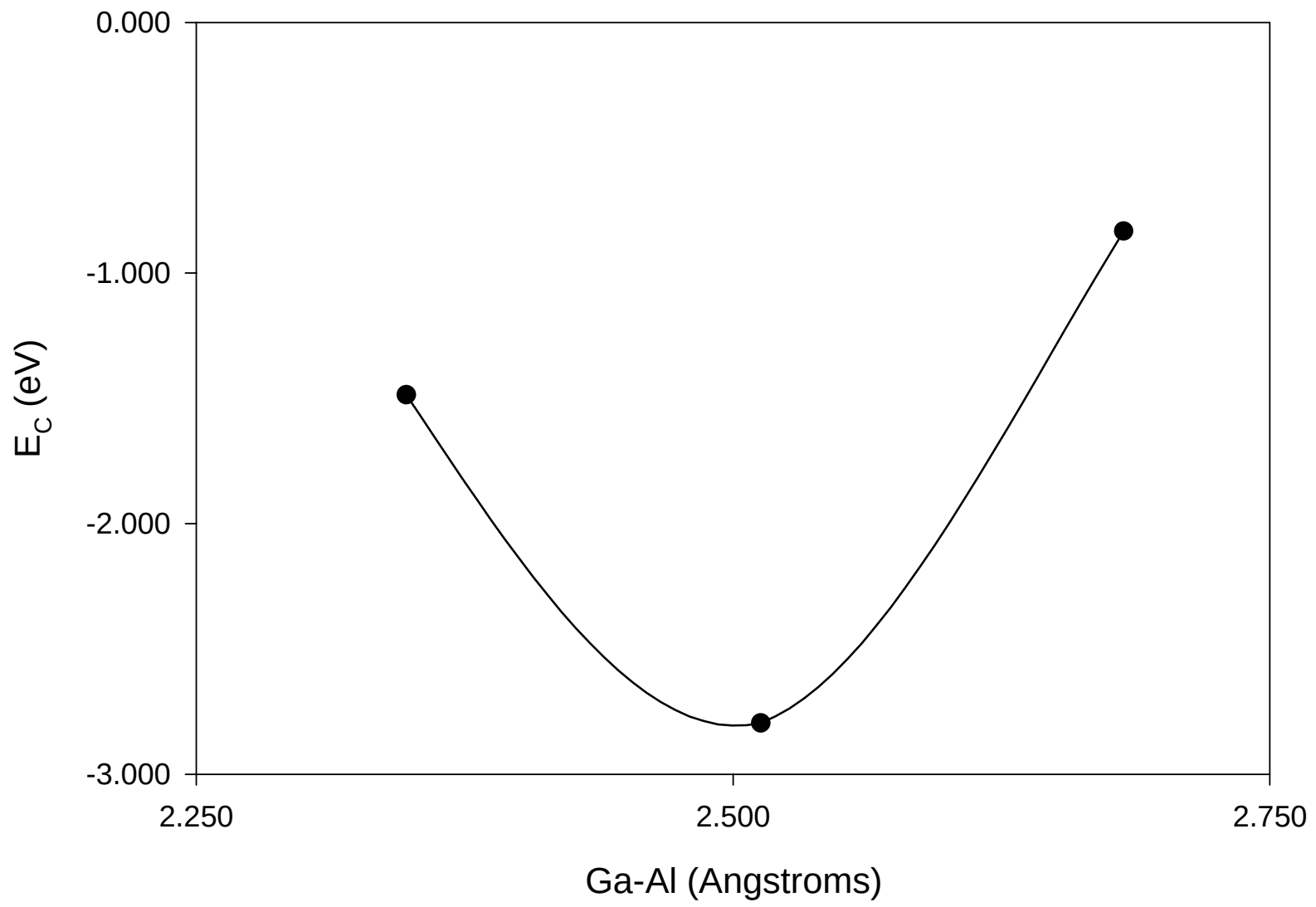

Fig. 23. Chemisorption energy vs. nearest surface neighbor bond length for $Ga_7As_6H_{19}$ + Al bridge site 2 with Hay-Wadt pseudopotential on Al.

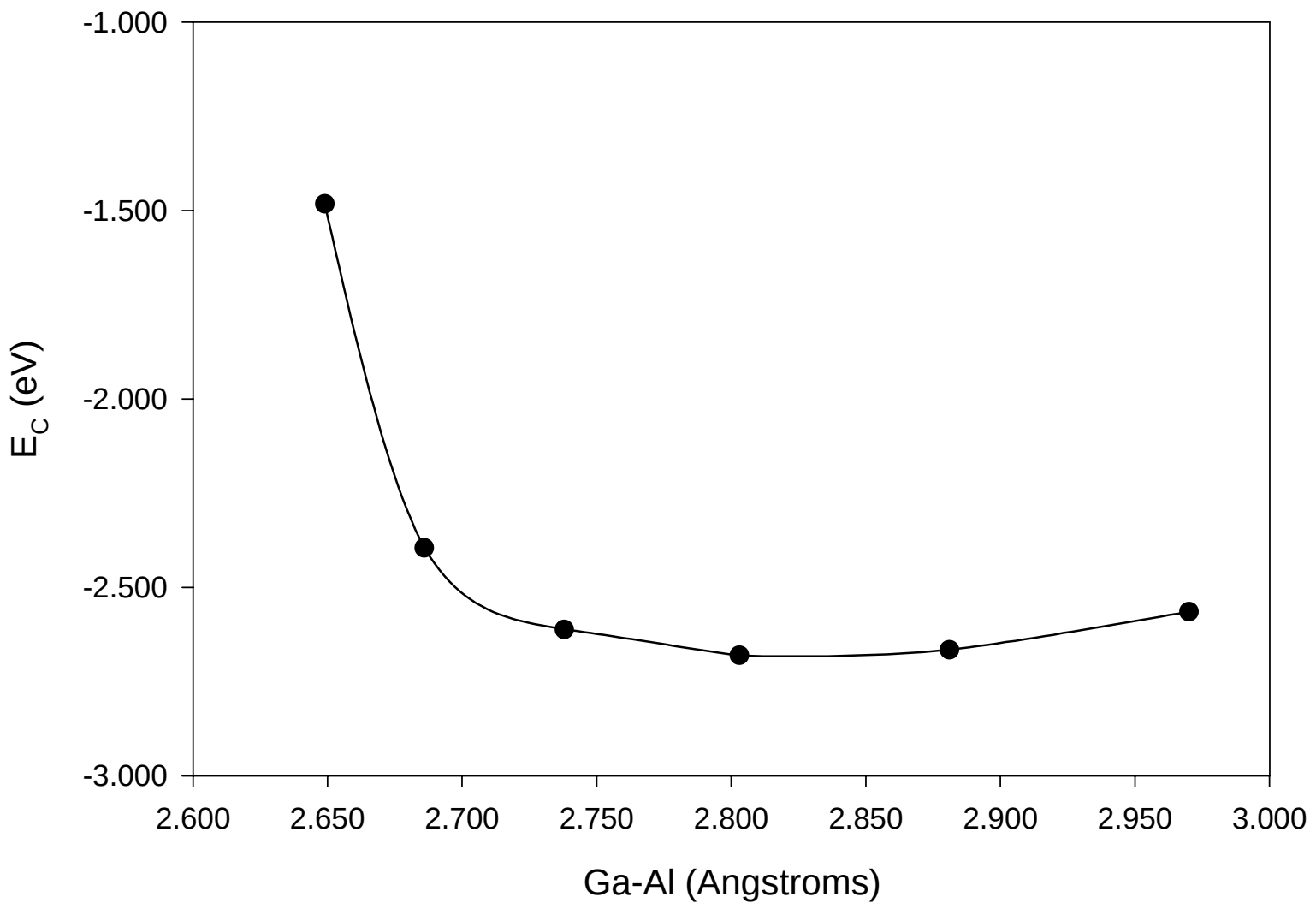

Fig. 24. Chemisorption energy vs. nearest surface neighbor bond length for $Ga_7As_6H_{19}$ + Al hollow site 3 with 6-311++G** basis on Al.

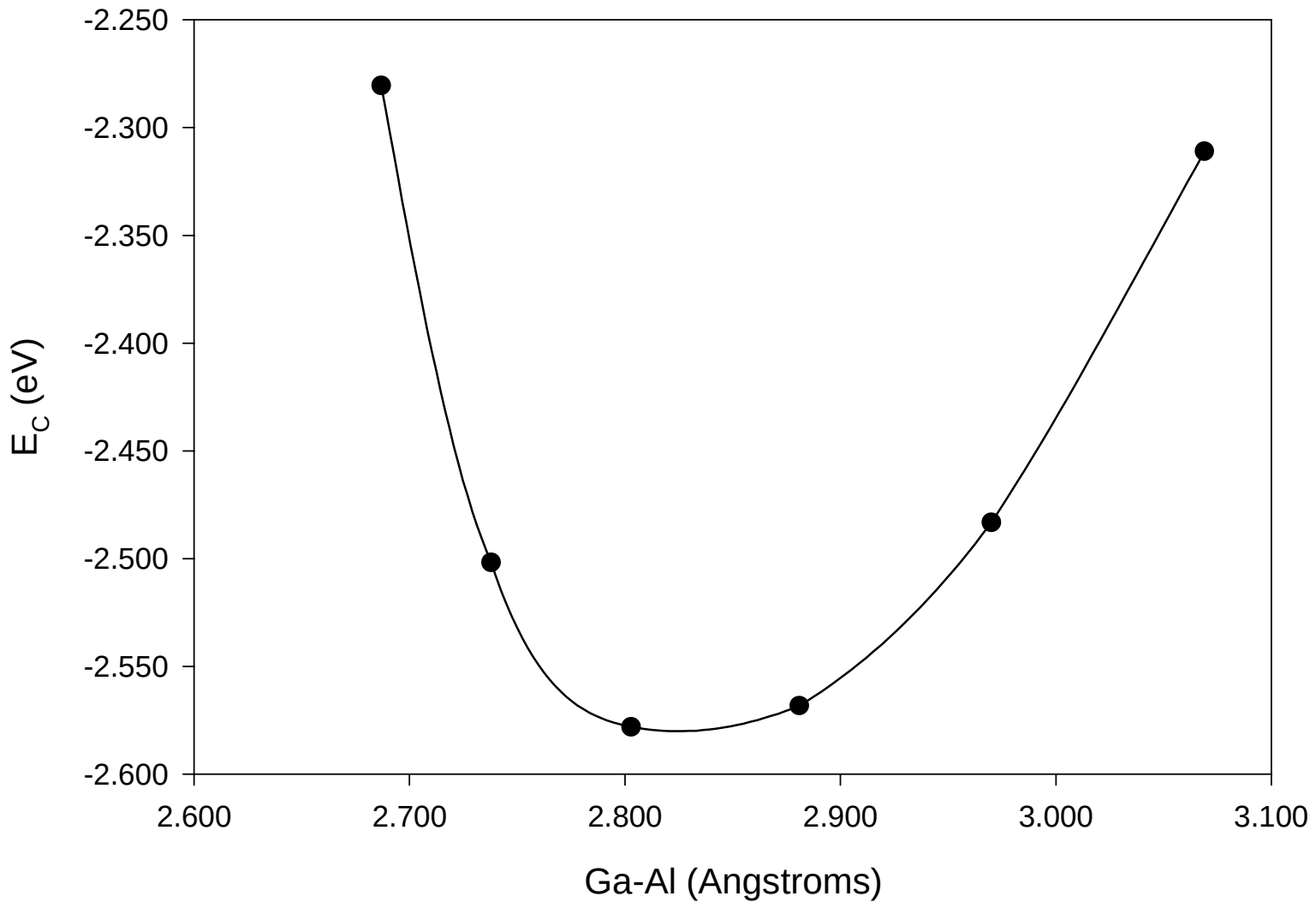

Fig. 25. Chemisorption energy vs. nearest surface neighbor bond length for $Ga_7As_6H_{19}$ + Al hollow site 3 with Hay-Wadt pseudopotential on Al.

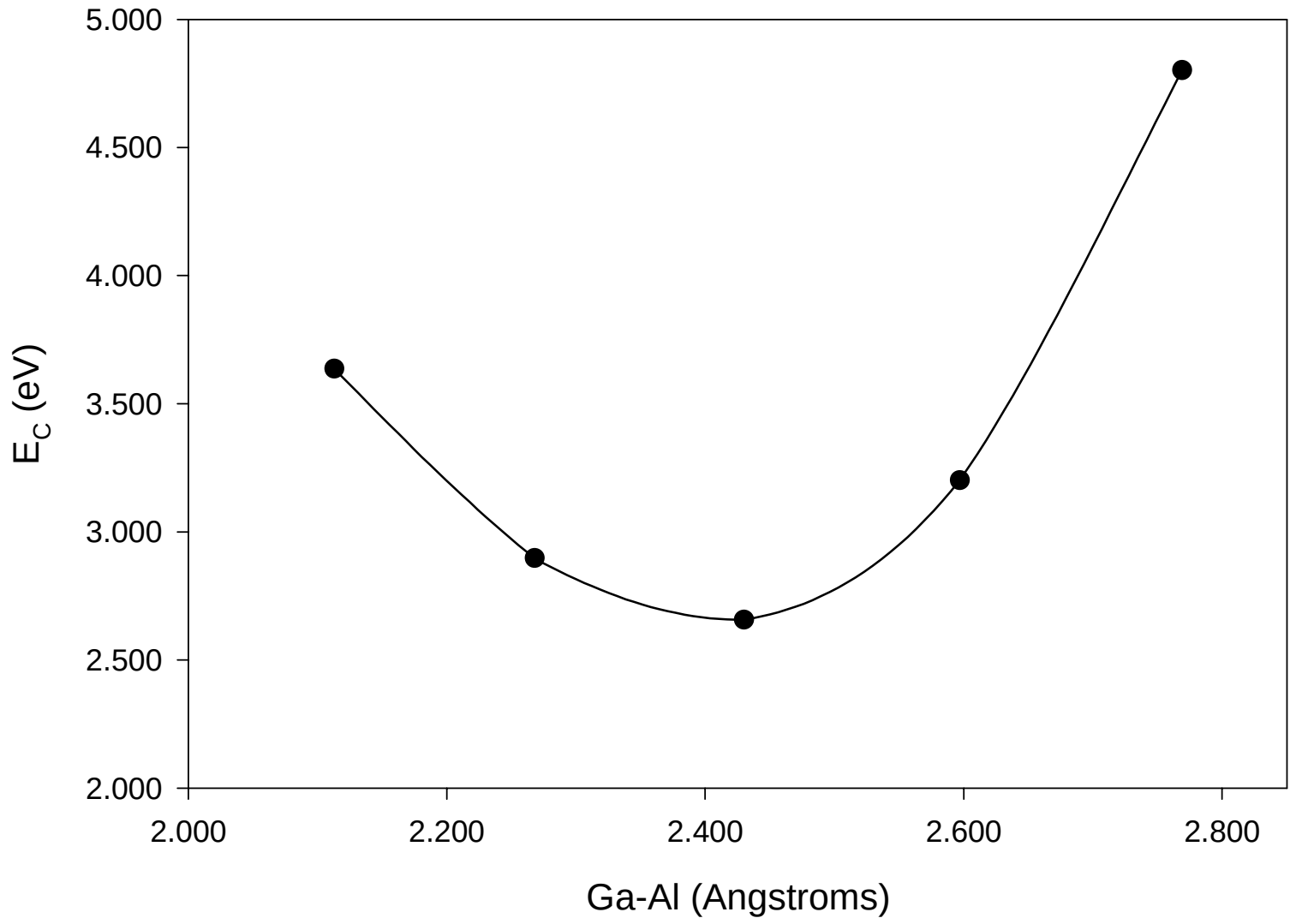

Fig. 26. Chemisorption energy vs. nearest surface neighbor bond length for $Ga_7As_6H_{19}$ + Al cage site 6 with 6-311++G** basis on Al.

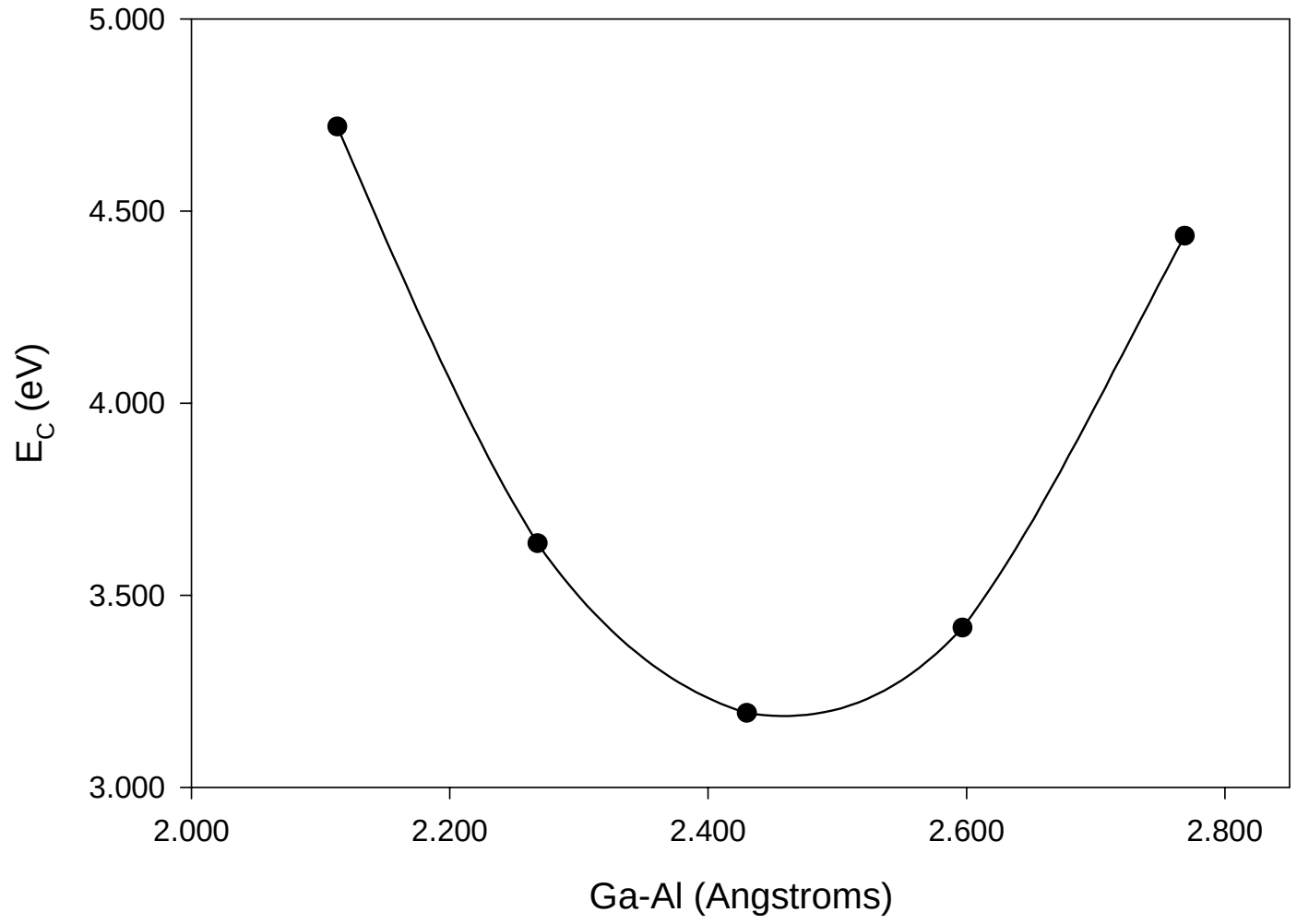

Fig. 27. Chemisorption energy vs. nearest surface neighbor bond length for $Ga_7As_6H_{19}$ + Al cage site 6 with Hay-Wadt pseudopotential on Al.

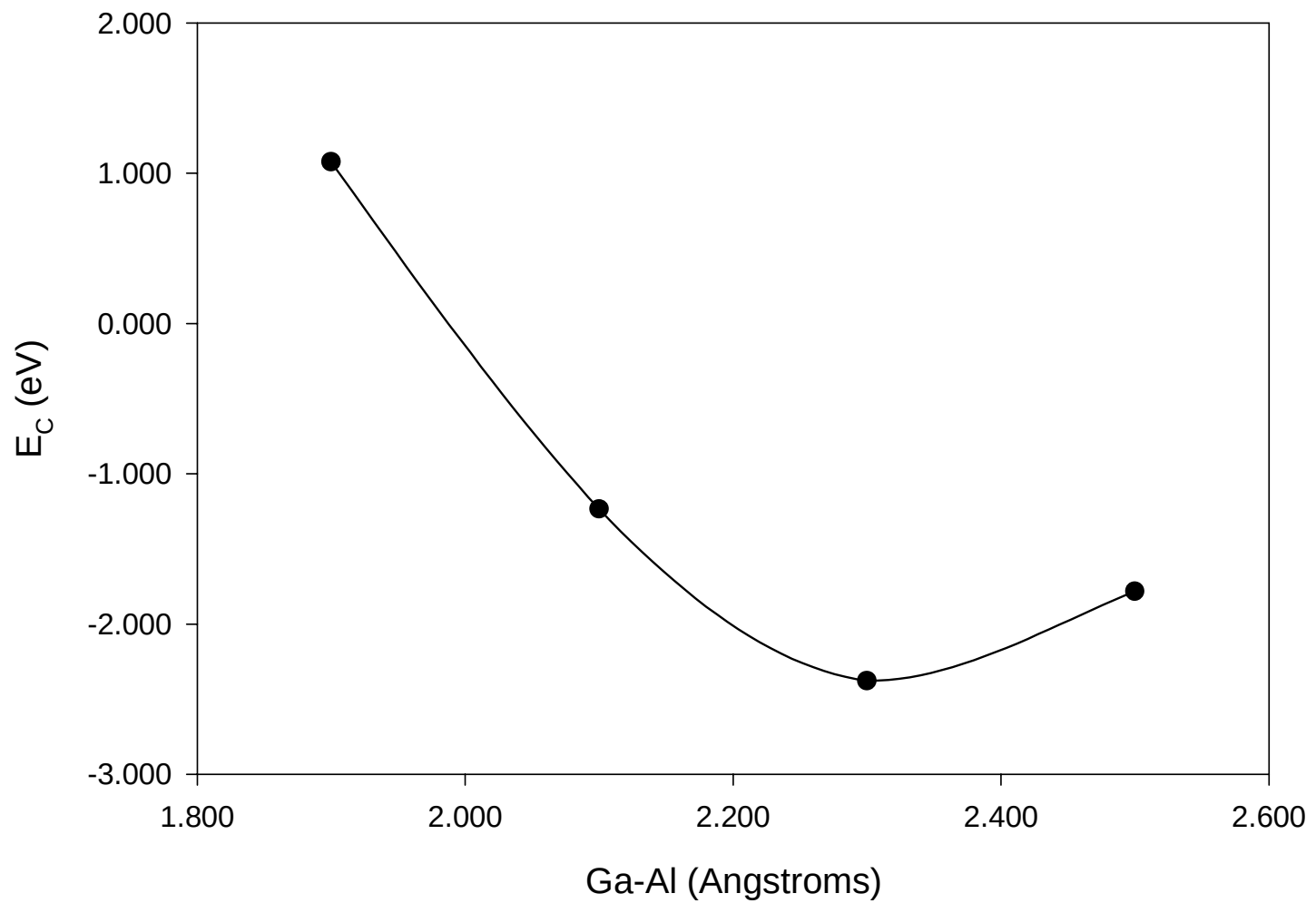

Fig. 28. Chemisorption energy vs. nearest surface neighbor bond length for $Ga_{19}As_{15}H_{39}$ + Al top site 1b with 6-311++G** basis on Al.

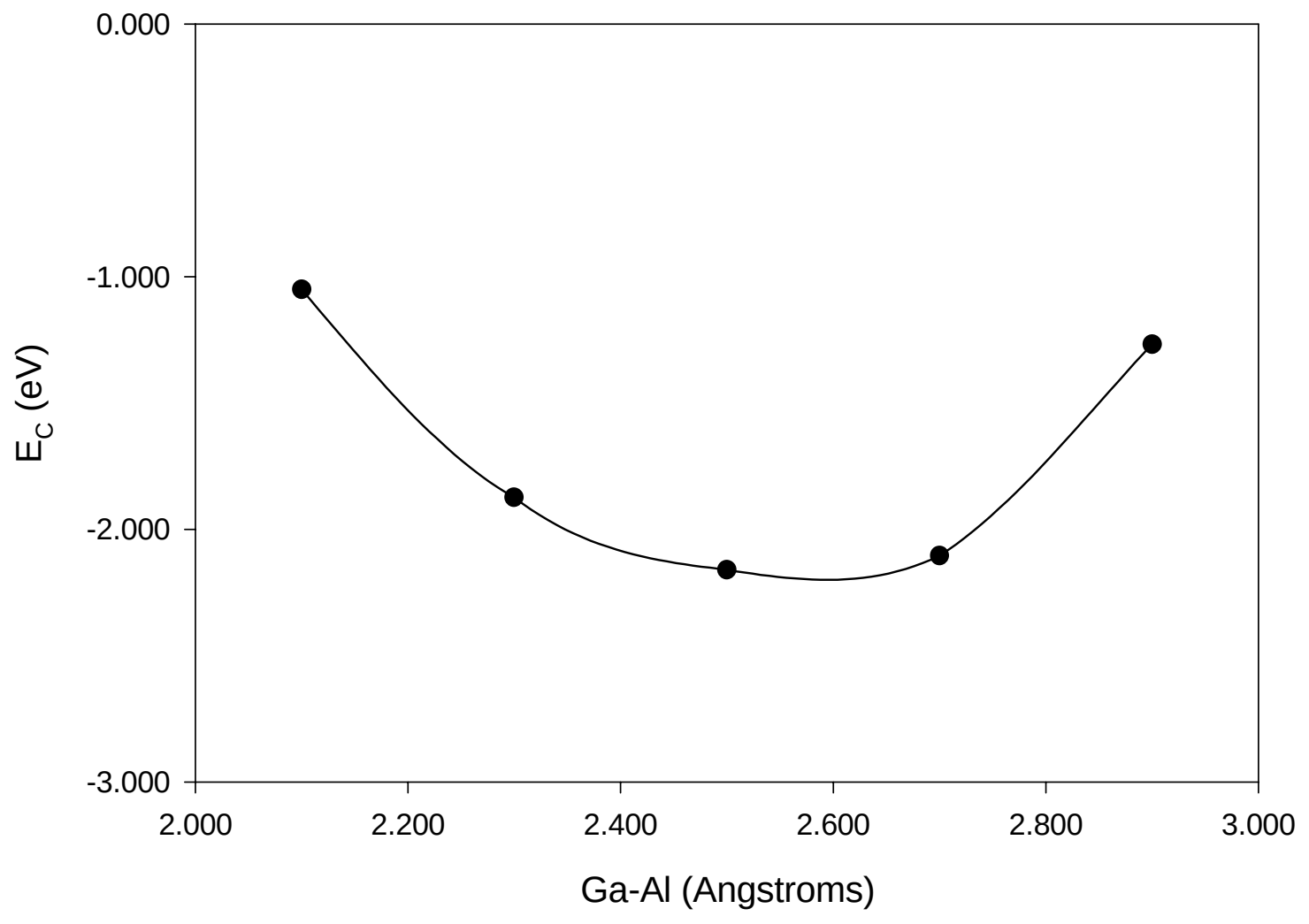

Fig. 29. Chemisorption energy vs. nearest surface neighbor bond length for $Ga_{19}As_{15}H_{39}$ + Al top site 1b with Hay-Wadt pseudopotential on Al.

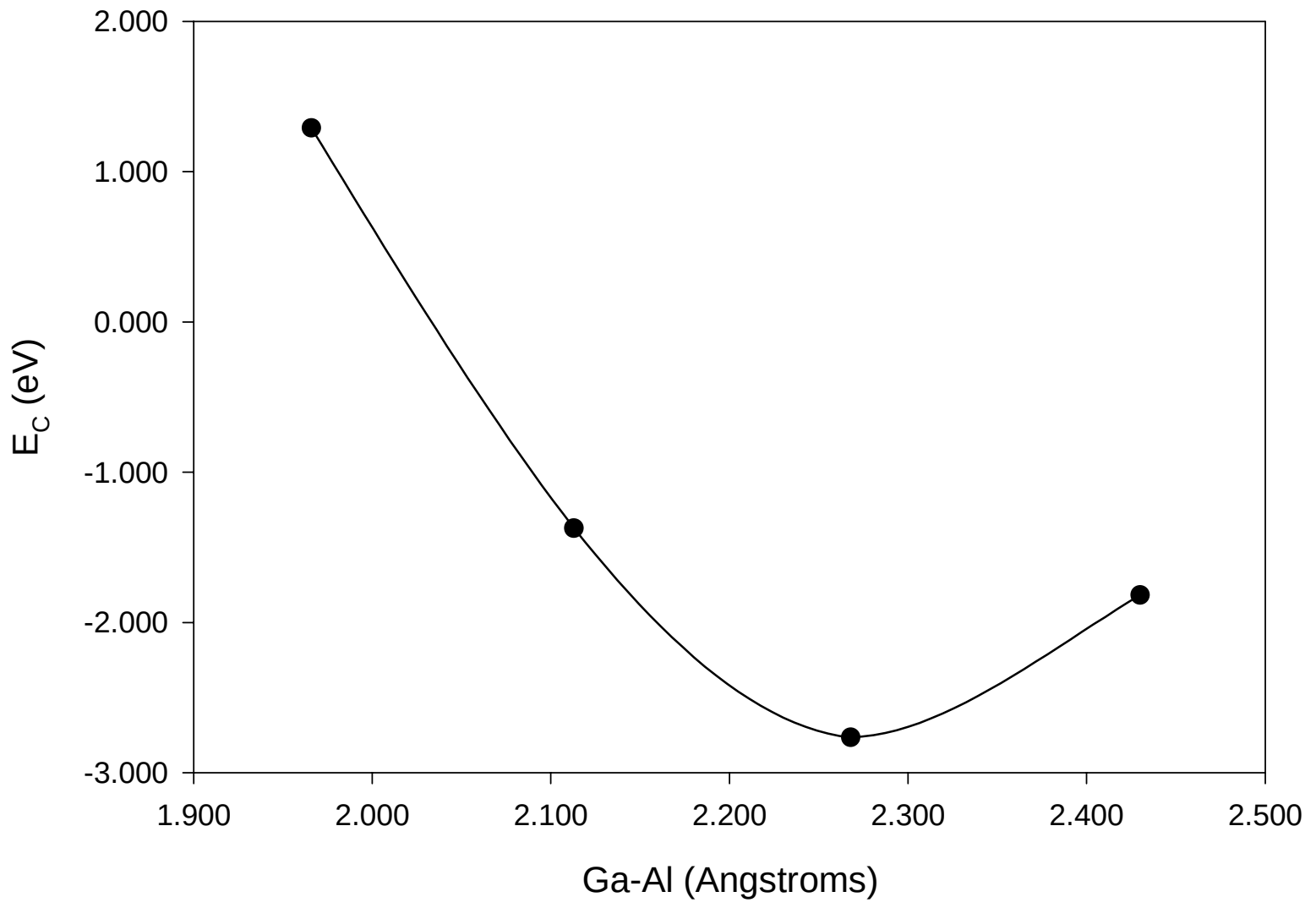

Fig. 30. Chemisorption energy vs. nearest surface neighbor bond length for $Ga_{19}As_{15}H_{39}$ + Al bridge site 2 with 6-311++G** basis on Al.

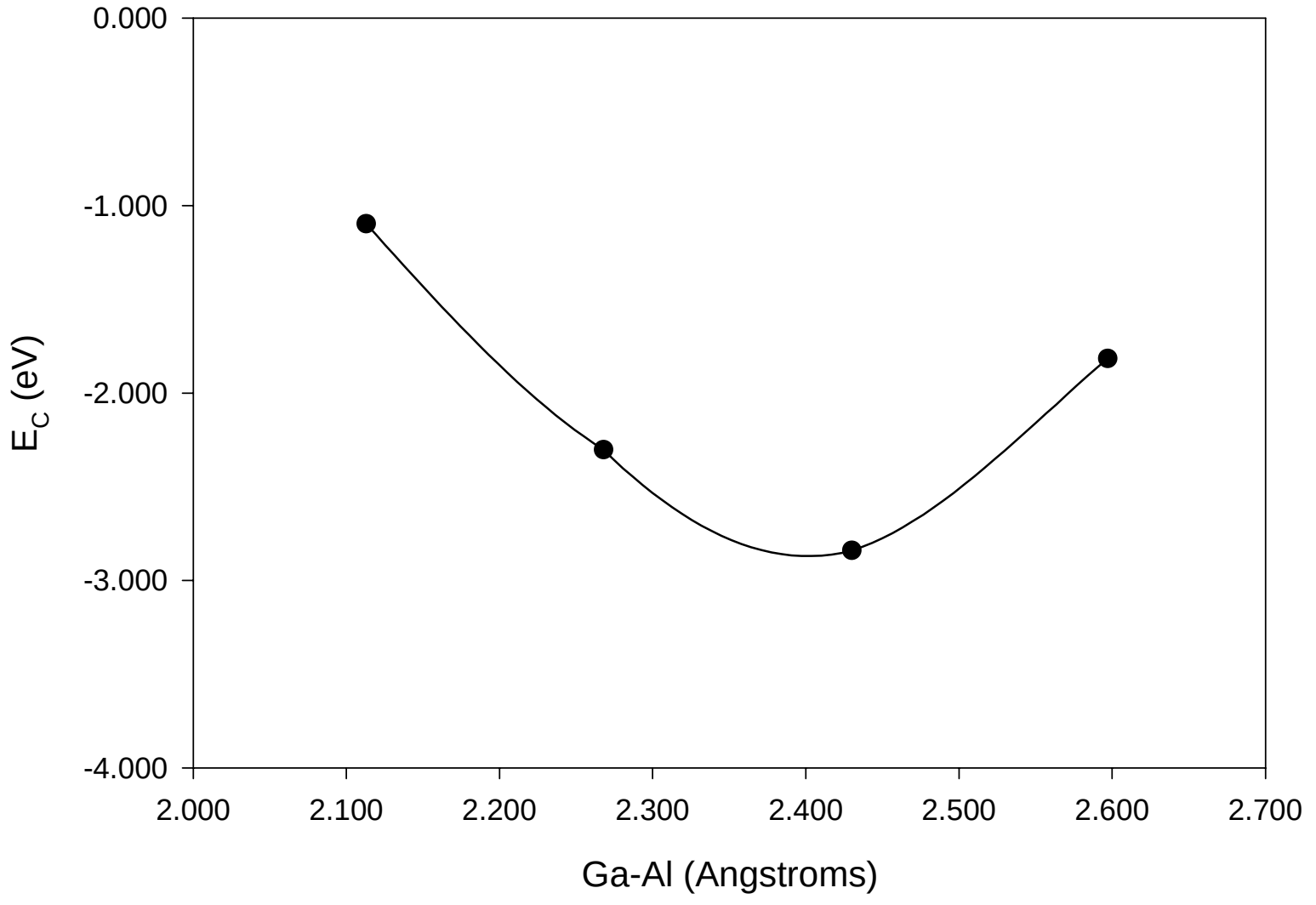

Fig. 31. Chemisorption energy vs. nearest surface neighbor bond length for $Ga_{19}As_{15}H_{39}$ + Al bridge site 2 with Hay-Wadt pseudopotential on Al.

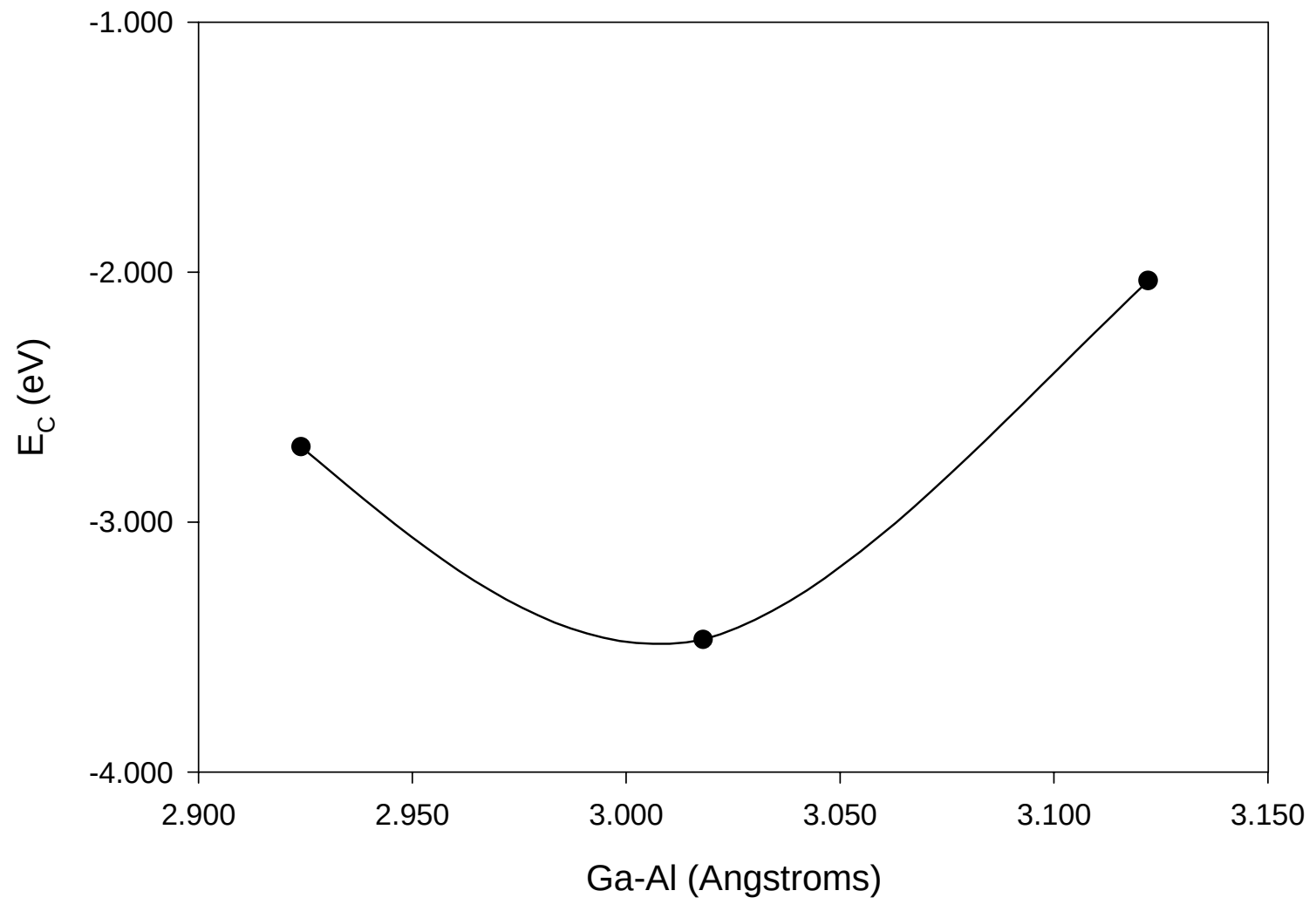

Fig. 32. Chemisorption energy vs. nearest surface neighbor bond length for $Ga_{19}As_{15}H_{39}$ + Al hollow site 3 with 6-311++G** basis on Al.

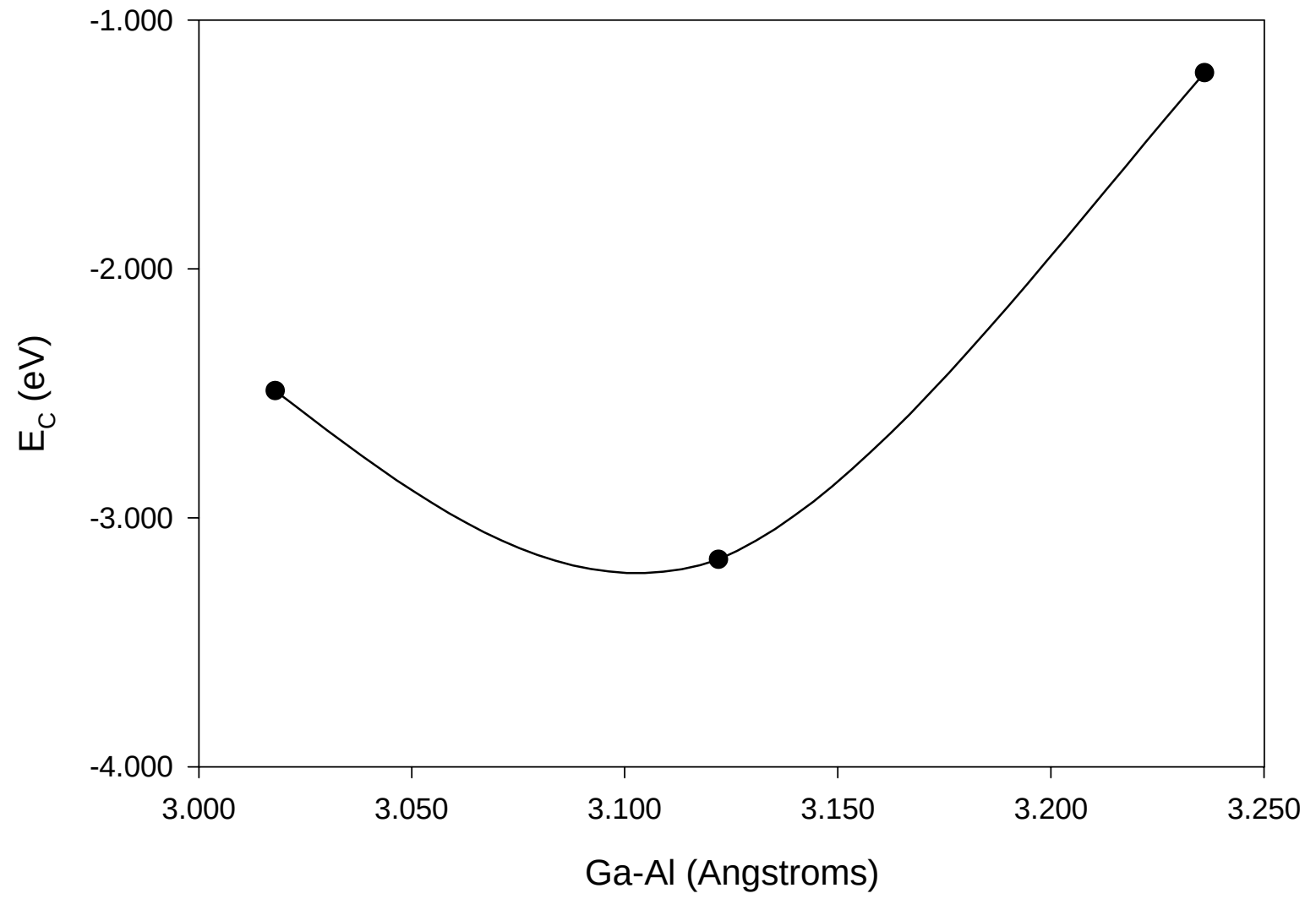

Fig. 33. Chemisorption energy vs. nearest surface neighbor bond length for $Ga_{19}As_{15}H_{39}$ + Al hollow site 3 with Hay-Wadt pseudopotential on Al.

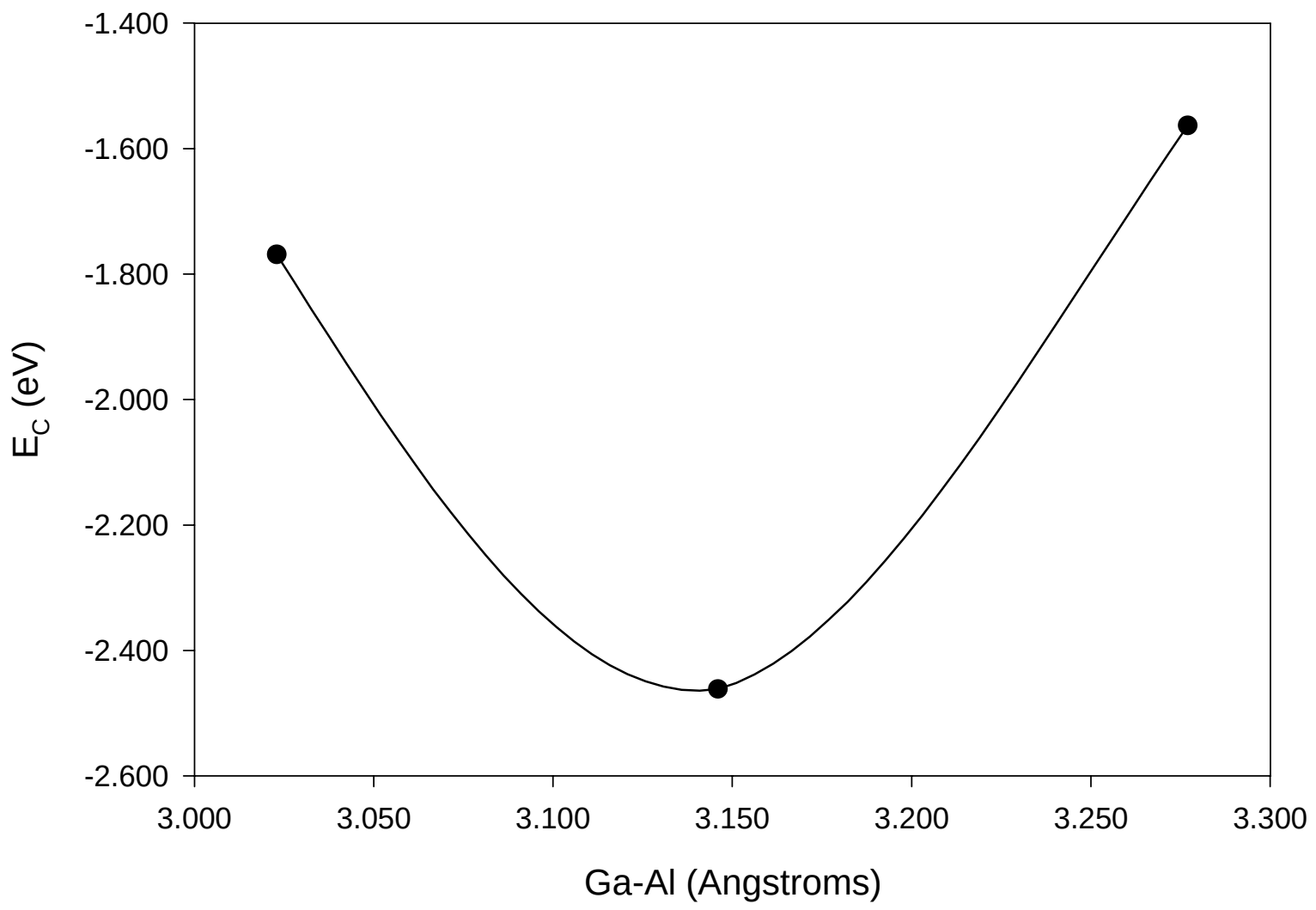

Fig. 34. Chemisorption energy vs. nearest surface neighbor bond length for $Ga_{19}As_{15}H_{39}$ + Al cave site 4 with 6-311++G** basis on Al.

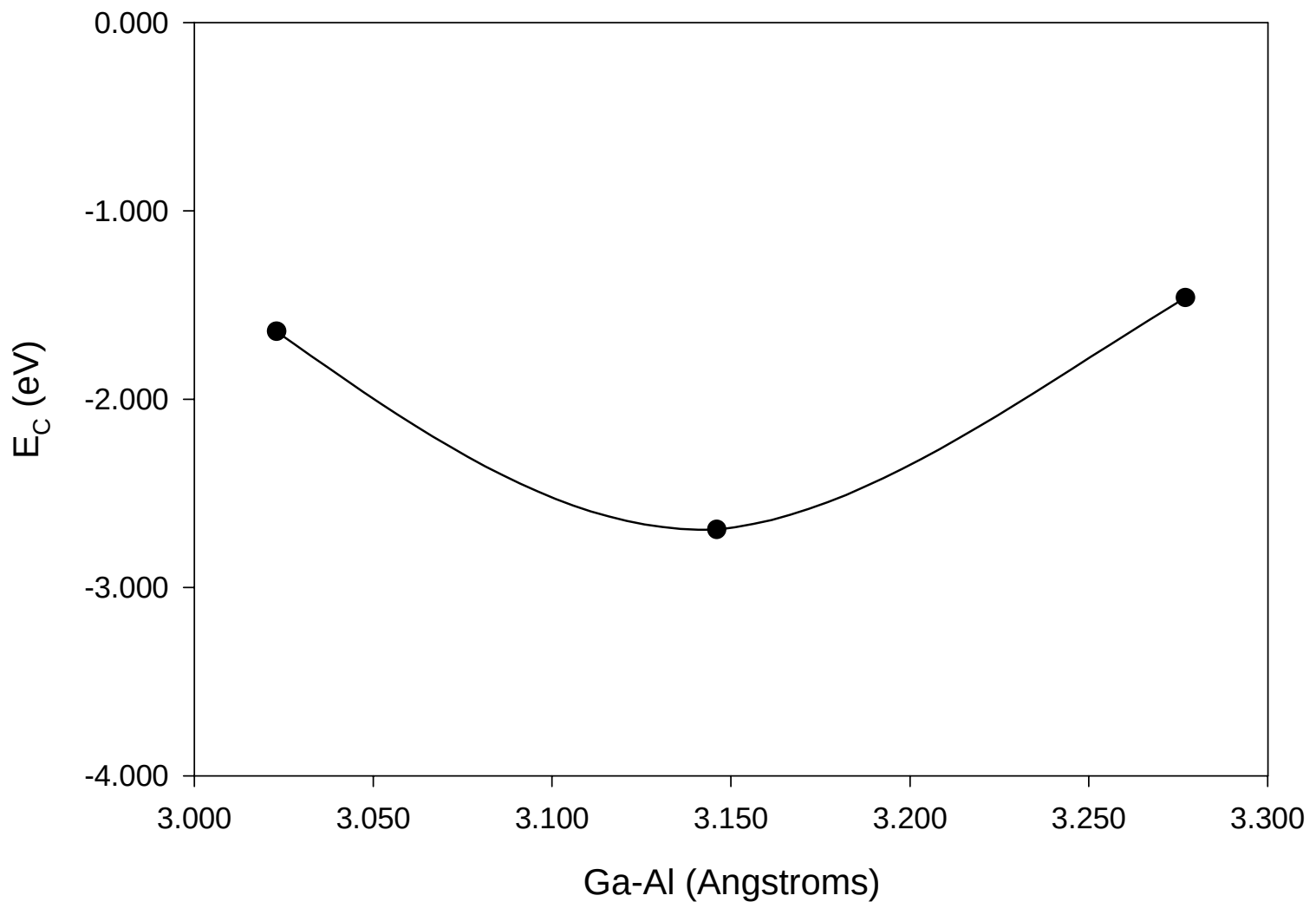

Fig. 35. Chemisorption energy vs. nearest surface neighbor bond length for $Ga_{19}As_{15}H_{39}$ + Al cave site 4 with Hay-Wadt pseudopotential on Al.

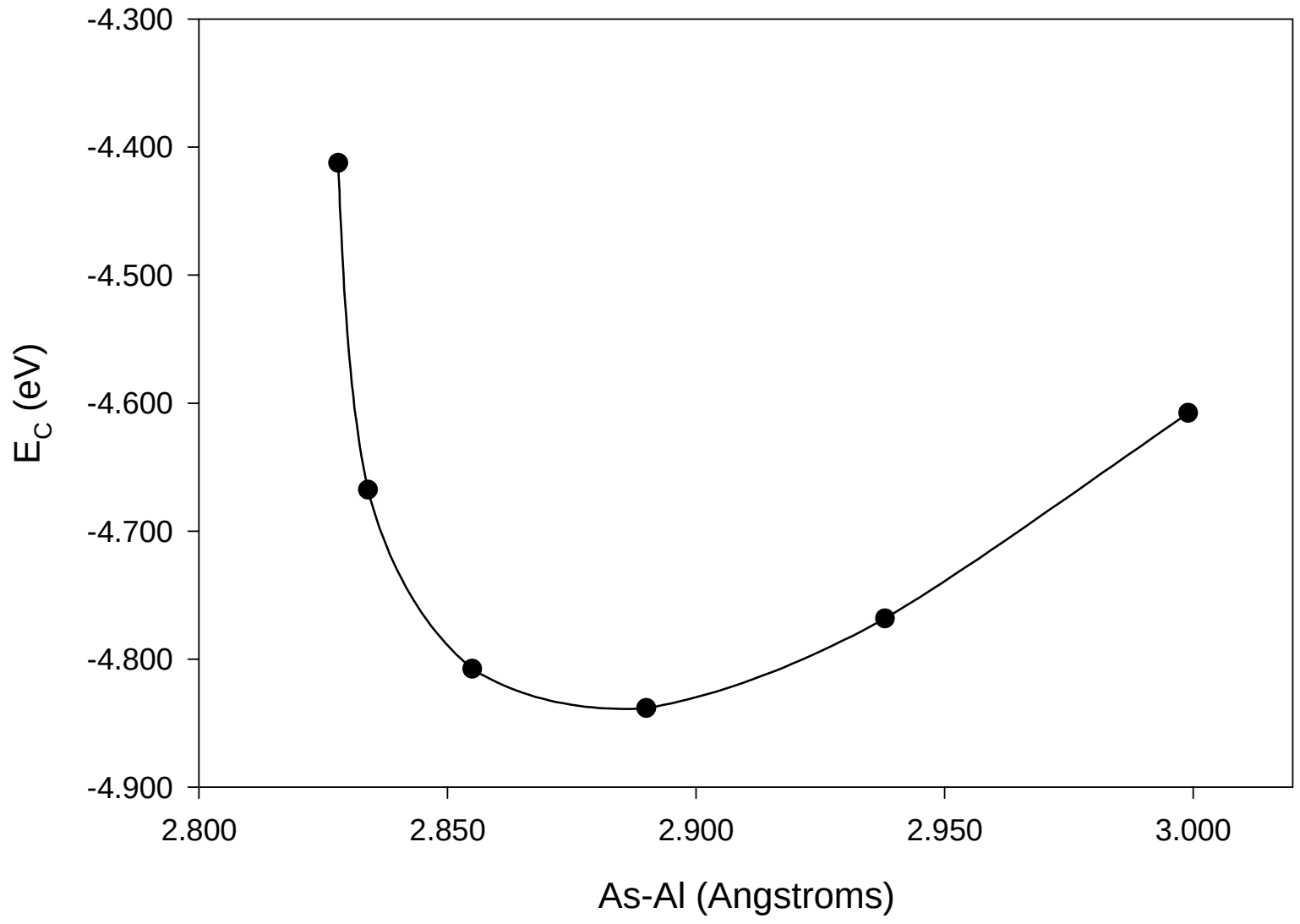

Fig. 36. Chemisorption energy vs. nearest surface neighbor bond length for $Ga_{19}As_{15}H_{39}$ + Al trough site 5a with 6-311++G** basis on Al.

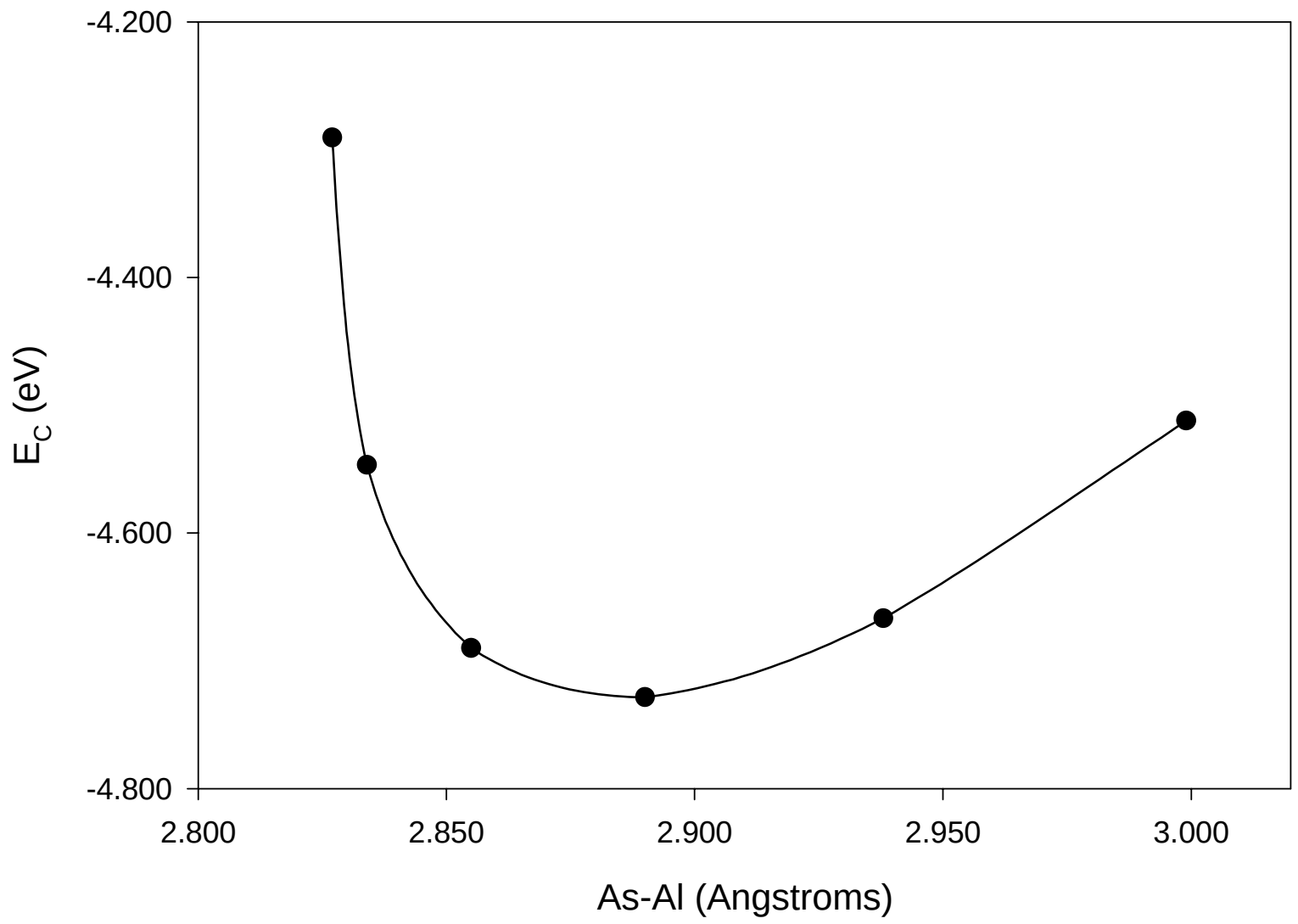

Fig. 37. Chemisorption energy vs. nearest surface neighbor bond length for $Ga_{19}As_{15}H_{39}$ + Al trough site 5a with Hay-Wadt pseudopotential on Al.

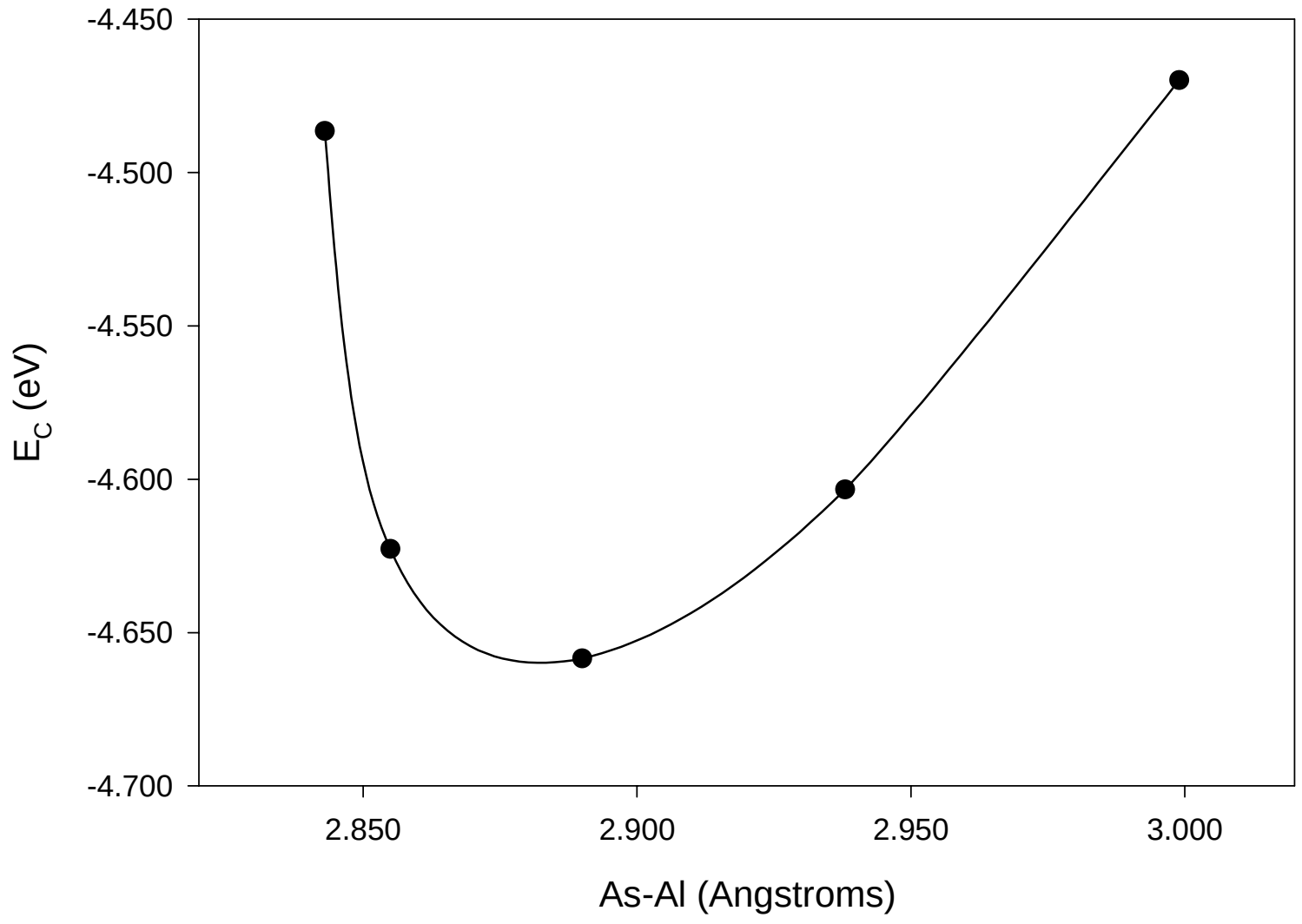

Fig. 38. Chemisorption energy vs. nearest surface neighbor bond length for $Ga_{19}As_{15}H_{39}$ + Al trough site 5b with 6-311++G** basis on Al.

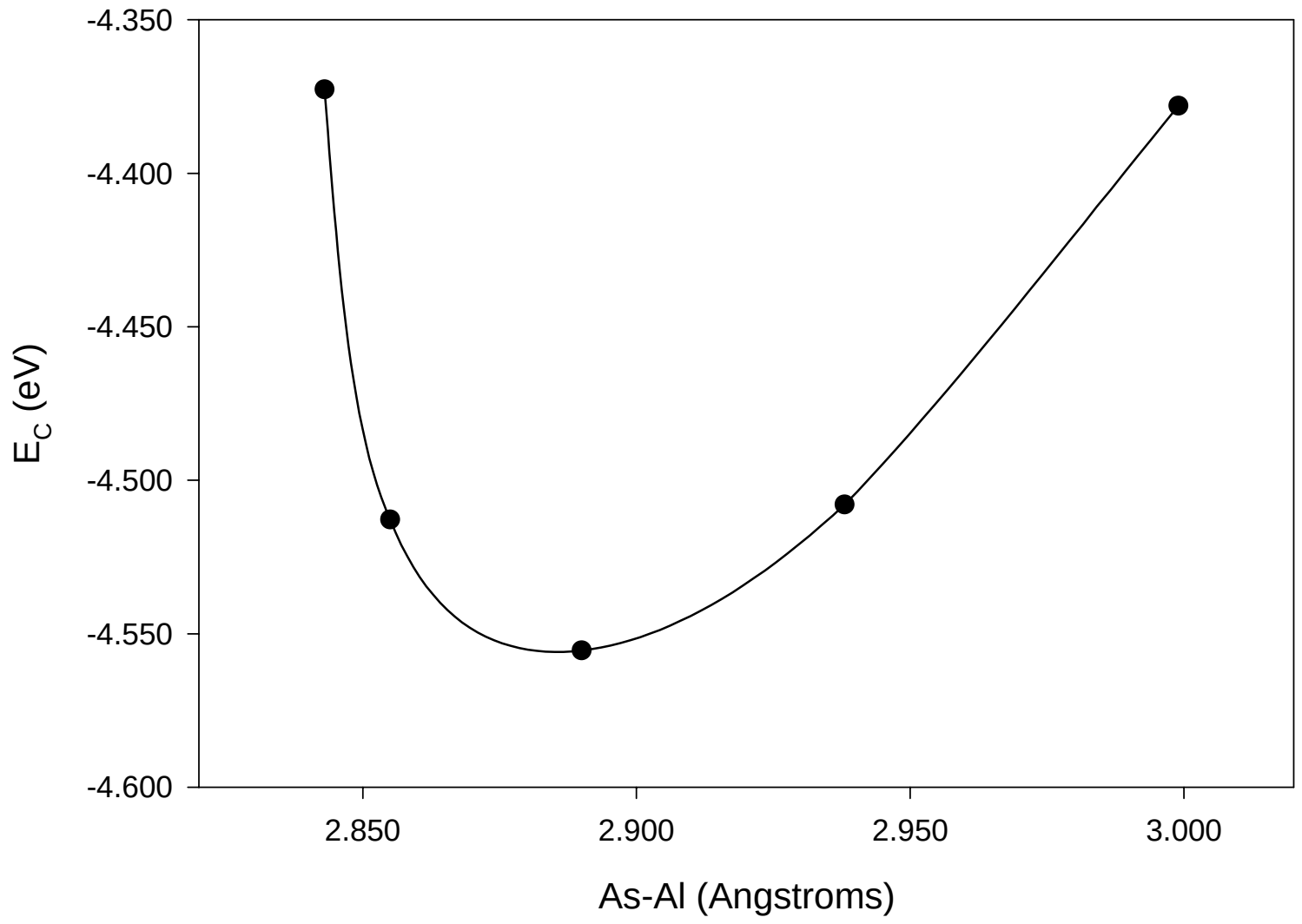

Fig. 39. Chemisorption energy vs. nearest surface neighbor bond length for $Ga_{19}As_{15}H_{39}$ + Al trough site 5b with Hay-Wadt pseudopotential on Al.